\documentclass[preprint,3p,times,twocolumn,authoryear]{elsarticle}

\usepackage{amsmath}
\usepackage{mathrsfs}
\usepackage{graphicx}
\usepackage{subcaption}
\usepackage{bm}
\usepackage[dvipsnames]{xcolor}
\journal{Control Engineering Practice}

\begin{document}

\graphicspath{{./Figures/}}

\begin{frontmatter}

\title{Data-Driven Repetitive Control: Wind Tunnel Experiments under Turbulent Conditions}

\author[DCSC]{Joeri Frederik\corref{cor1}}
\ead{j.a.frederik@tudelft.nl}
\cortext[cor1]{Corresponding author}
\author[ForWind]{Lars Kr{\"o}ger}
\author[ForWind]{Gerd G{\"u}lker}
\author[DCSC]{Jan-Willem van Wingerden}

\address[DCSC]{Delft Center of Systems \& Control (DCSC), Department of Mechanical, Maritime and Materials Engineering (3mE), Delft University of Technology, Mekelweg 2, 2628CD Delft, The Netherlands}
\address[ForWind]{ForWind - Institute of Physics, University of Oldenburg, 26111 Oldenburg, Germany}

\begin{abstract}
A commonly applied method to reduce the cost of wind energy, is alleviating the periodic loads on turbine blades using Individual Pitch Control (IPC). In this paper, a data-driven IPC methodology called Subspace Predictive Repetitive Control (SPRC) is employed. The effectiveness of SPRC will be demonstrated on a scaled 2-bladed wind turbine. An open-jet wind tunnel with an innovative active grid is employed to generate reproducible turbulent wind conditions. A significant load reduction  with limited actuator duty is achieved even under these high turbulent conditions. Furthermore, it will be demonstrated that SPRC is able to adapt to changing operating conditions.
\end{abstract}


\end{frontmatter}


\section{Introduction}
\label{sec:intro}

In the quest to make the cost of wind energy increasingly competitive with conventional energy sources such as fossil fuels, wind turbine structures become increasingly larger and more slender in order to increase their rated power \citep{vkuik}. Consequently, the loads experienced by the blades of turbines also increase, and it becomes of vital importance to mitigate these loads. 

The majority of dynamic loads on wind turbine rotors have a periodic nature, caused by wind shear, tower shadow, gravity and partial wake overlap from upwind turbines \citep{ipc1}. To reduce these deterministic loads, Individual Pitch Control (IPC) is a method receiving an increasing amount of attention \citep{review}. In IPC, the pitch angle of each blade is, as the name suggests, controlled individually to decrease the out-of-plane bending moments. This method is relatively easy to implement, since modern wind turbines already have individual pitch capabilities, as well as measurements of the bending moments. By applying periodic pitch angles to the blades on top of the collective pitch, significant load alleviations can be achieved \citep{ipc1}.

Many different IPC approaches are studied in literature. Initially, the focus was mainly on controlling the load occuring once per rotation (1P) using Linear Quadratic Gaussian (LQG) controllers to solve the multiple-input multiple-output (MIMO) problem \citep{lqg1,ipc3}. However, since the 1P loads are symmetric, these loads do not cause the largest loads on the non-rotating parts of the wind turbine structure. These parts experience the largest loads at the blade passing frequency $N$P, with $N$ the number of blades of the turbine, \citep{ipc2}. One method of alleviating these $N$P loads is by applying the Coleman transformation \citep{coleman}, which transforms the loads into a static reference frame. This allows the use of simple linear single-input single-output (SISO) control methods, such as PI-controllers \citep{ipc2,edwin2}. 

An important downside of IPC is the substantial increase of the pitch actuator duty cycle. Subsequently, the wear on the bearings of the blades is also increased. In the proposed IPC methods, this effect could be enlarged at higher wind turbulence intensities, as these methods might attempt to also control the non-deterministic loads. However, this is a research area that has not yet received a lot of attention. Furthermore, the mentioned IPC algorithms assume a constant operating conditions, and are usually not able to adapt to changing rotor velocities. 

A novel IPC methodology that deals with both these problems is proposed in \cite{SPRC1}. This methodology called Subspace Predictive Repetitive Control (SPRC) and combines subspace identification \citep{vdveen2} with repetitive control. By using measurement data to do online identification, the model can be refined during operation. Furthermore, the repetitive control law targets only the specified deterministic loads, thus lowering the actuator duty cycle. SPRC shows promising results in simulations \citep{SPRC1} and in wind tunnel experiments with laminar flow conditions \citep{SPRC2}. These laminar flow conditions are however not a realistic representation of the wind conditions that a turbine in the field would experience. 

\begin{figure}[!b]
\centering
\includegraphics[width=0.48\textwidth]{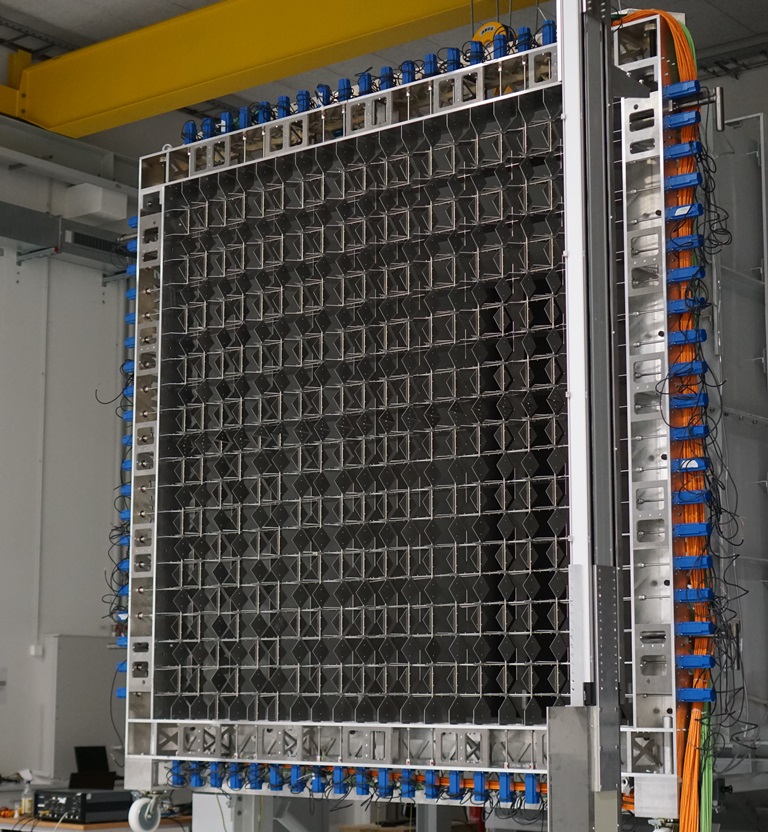}
\caption{The active grid mounted on the $3 \times 3\,$m wind tunnel inlet in open test section configuration.}
\label{fig:grid}
\end{figure}

In this paper, experiments will be presented that form the next vital step in assessing the relevance of SPRC as an IPC algorithm. Using the open jet wind tunnel of ForWind at the University of Oldenburg, which is equipped with a novel active grid, realistic turbulent wind conditions can be created. Furthermore, the active grid makes it possible to reproduce these conditions, thus enabling an evaluation of different control methodologies. Preliminary results of these experiments are shown in \cite{ikke}. Modifications to the SPRC algorithm are proposed that make it possible to achieve superior performance compared to other IPC methodologies. Compared to \cite{ikke},  results showing the capabilities of SPRC to adapt to changing operating conditions in a turbulent flow field will be presented.

The structure of this paper is as follows: in Section~\ref{sec:setup}, the experimental setup is described. This section contains a description of the flow conditions as created by the active grid (\ref{sec:grid}), a description of the wind turbine (\ref{sec:turbine}), and an overview of the real-time environment (\ref{sec:rt}). Section \ref{sec:sprc} covers the SPRC algorithm and its modifications, and Section \ref{sec:results} will then show the results of this algorithm subject to turbulent wind conditions. Finally, conclusions will be drawn in Section \ref{sec:conclusions}.

\section{Test Setup}\label{sec:setup}

In this section, the test setup used to conduct the experiments is described. First, the wind tunnel equiped with the novel active grid will be explained, followed by a description of the two-bladed control-oriented wind turbine. Finally, an overview of the real-time environment will be given.

\subsection{Active Grid}\label{sec:grid}

The experiments shown in this paper have been conducted in a low-speed wind tunnel of the University of Oldenburg. This tunnel has a cross section of $3 \times 3\,$m and can reach wind speeds up to 30$\,$m/s. On the inlet of this tunnel, an active grid is mounted as shown in Figure \ref{fig:grid}. This active grid consists of 20 servomotors at each side that are connected to an axis mounted with rigid square flaps, as introduced by \cite{makita}. Consequently, the 80 different axes of the active grid can be actuated individually. The change of the angle $\gamma$ of the rigid square flaps with respect to the inflowing wind results in either a blockage or a deflection of the inflow. 

By dynamically varying $\gamma$ over time, various turbulent flow fields with specific characteristics such as atmospheric turbulence can be generated at certain positions in the test section \citep{knebel,heisselmann}. A comprehensive overview of the work in active grid research can be found in the review article of \cite{Mydlarski}. By repeating a predefined dynamic sequence of input angles $\gamma$, defined as an excitation protocol, it is possible to accurately reproduce turbulent flow fields.  

\begin{figure}[t]
\centering
\includegraphics[width=0.48\textwidth]{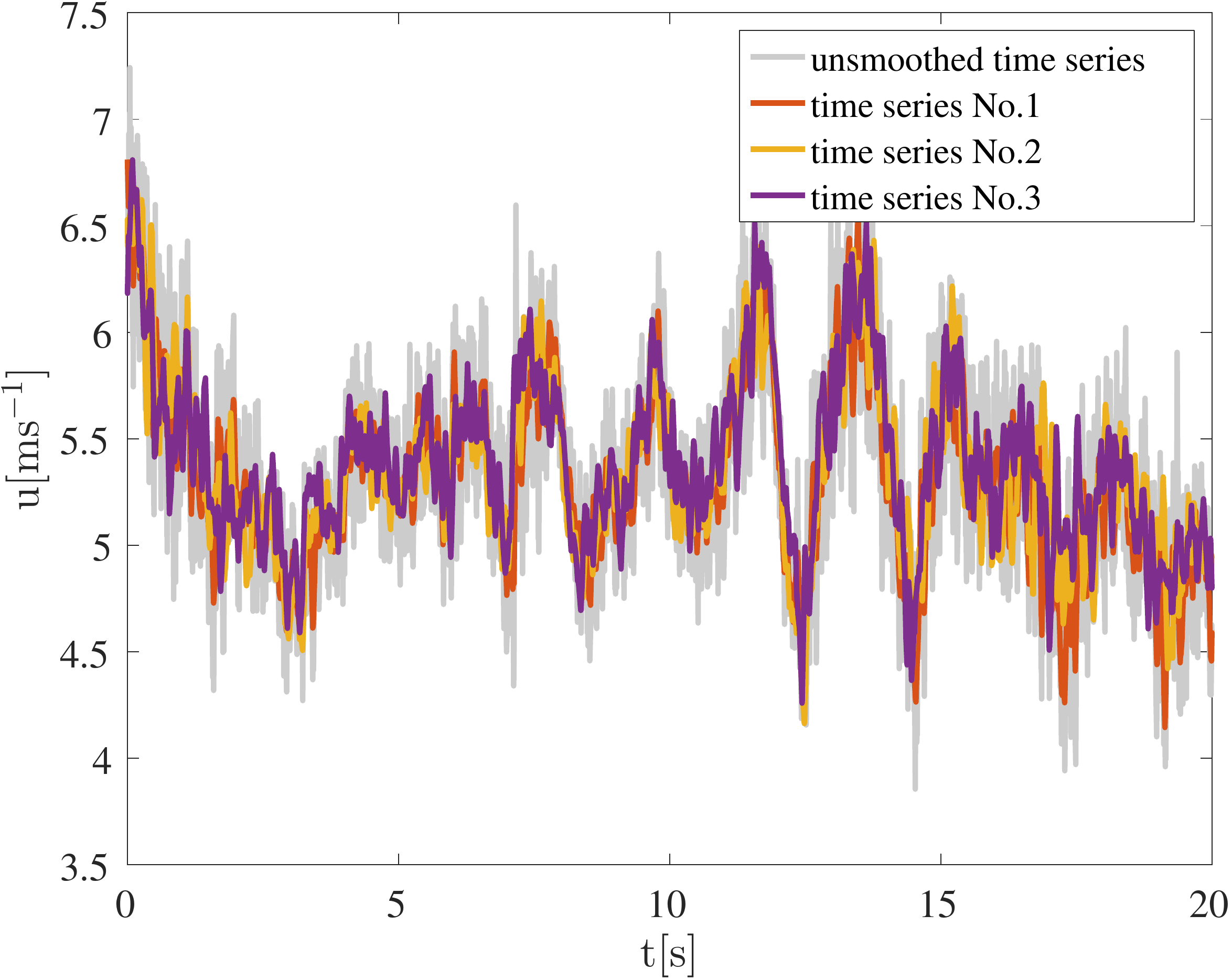}
\caption{Three exemplary wind speed time series generated by the Lidar excitation protocol smoothed by a moving average filter for a better comparison. An unfiltered wind speed time series is shown as reference in light grey.}
\label{fig:timeseries}
\end{figure}

To validate the new control concepts of the model wind turbine in turbulent conditions and to validate the reproducibility of the inflow, the flow field acting on the wind turbine is characterized. This was realized using a 2D hotwire system by Dantec Dynamics. This sensor consists of a thin wire suspended between two prongs and measures the wind speed and direction. An x-wire of the type 55P51 was used and operated at a sampling rate of 20$\,$kHz with a low-pass filter at 10$\,$kHz. For the data acquisition an 18-bit National Instruments A/D converter was used. These sensors were used to measure the wind speed at the location of the hub of the model turbine at 20 mesh sizes ($3\,$m) distance to the active grid. Additionally, the hotwire was shifted 1$\,$m to either side to determine differences in the flow field in the range of the wind turbine diameter.

Using the active grid described above, different wind conditions can be created. In this paper, the active grid was used in four different modes: two static and two active cases. For the static cases, the angle of attack of the active grid flaps was set to a constant angle of 0$^{\circ}$, corresponding to the orientation of the flaps with minimal blockage, and 45$^{\circ}$. In the active cases, two excitation protocols were used. The first one, called the Lidar mode, is based on Lidar measured atmospheric wind data and creates a wind field with intermittent behavior. The second one, called the gusts mode, is creating a mexican hat-like wind field with single gusts. 

The flow fields of all these modes were investigated for three different mean wind velocities of 4$\,$m/s, 4.5$\,$m/s and 5$\,$m/s. In the following, the different protocols are characterised briefly for the 5$\,$m/s test cases, in terms of reproducibility, flow characteristics, turbulence intensity (TI) and the dynamics in the power spectra. A full characterization including suplementary measurements shifted to the outer radius of the wind turbine and further analysis to determine the reproducibility and intermittency of the flow fields is presented in \cite{kroeger}. 

\begin{figure}[t]
\centering
\includegraphics[width=0.48\textwidth]{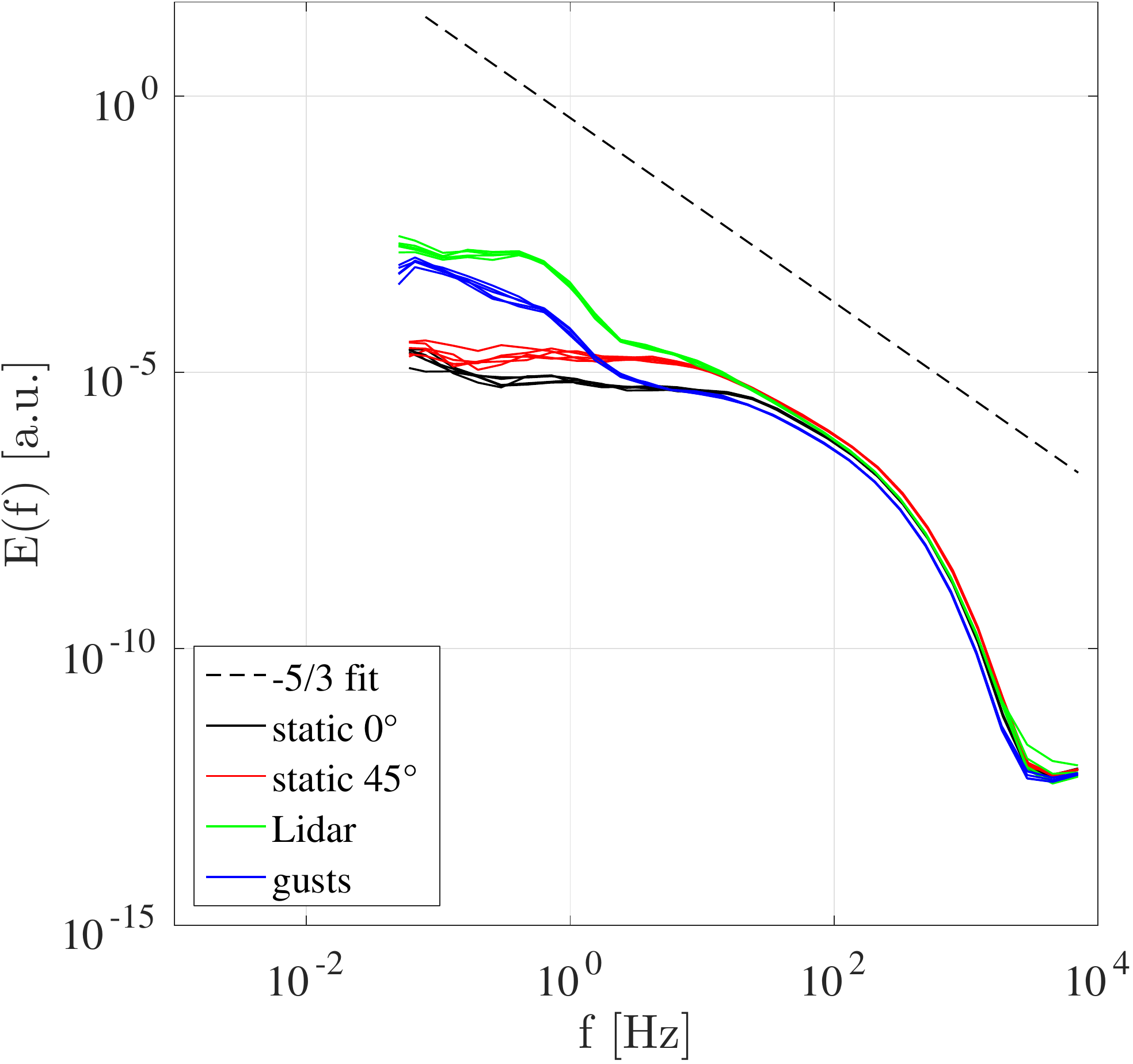}
\caption{Power spectra of all four active grid modes. For every mode the spectra of five repetitions are shown. The $-5/3$ law of the natural decay of turbulence is included as a reference (- - - -).}
\label{fig:FFT_all}
\end{figure}

To show the reproducibility, three wind speed time series are shown in Figure \ref{fig:timeseries} smoothed by a moving average filter using a subset of 1000 samples. In light grey, an unsmoothed time series is shown as a reference. These turbulent flow fields were generated by repeating the Lidar excitation protocol of the active grid. As shown, the main dynamic features in the flow are highly reproducible, whereas the higher-frequent components show differences in the direct comparison. By inspection of the smoothed power spectra of all four test cases shown in Figure \ref{fig:FFT_all}, further analysis of the dynamics of the flow fields is performed. To show the resemblance between the flow fields generated by repeated excitation, the spectra of five repeated time series are plotted on top of each other, appearing for all test cases nearly as a single line. 

As a reference, the $-5/3$ law of the natural decay of turbulence postulated by \cite{kolmogorov} is also shown, to compare the results with the theoretical values. This is valid for the higher frequency ranges of all test cases. The data of the two actively driven test cases both show a significant increase of the energy in the lower frequency ranges of $0.1-10\,$Hz, corresponding to structures in the flow with sizes of $0.5 - 50\,$m. This results in more realistic turbulent structures acting on the model wind turbine compared to using regular grids, with integral length scales in the range of their mesh width \citep{kroeger}.

To describe the variability of the wind field the turbulence intensity (TI) is used. The TI is defined as the ratio of the standard deviation and the mean of the wind speed time series: 
\begin{equation*}
\text{TI} = \frac{\sigma_{u}}{\langle u \rangle}.
\end{equation*}
As every test case was repeated five times, the mean value of the TI over these experiments were determined for the different modes and are shown in Table \ref{tab:modes}. Note that although the gusts mode has relatively low average TI's, this mode creates the largest variations in wind velocity. As a consequence, it might result in higher load variations on the blades than would be expected based on the TI.

The mean turbulence intensities (TI's) for the different modes are shown in Table \ref{tab:modes}. Note that although the gusts mode has relatively low average TI's, this mode creates the largest variations in wind velocity. As a result, it might result in higher load variations on the blades than would be expected based on the TI.

\begin{table}[h]
\caption{The average turbulence intensities for different modes of the active grid.}\label{tab:modes}
\begin{tabular}{ccc}
\hline
\textbf{Mode} & \textbf{Centerline TI [$\%$]} & \textbf{Shifted TI [$\%$]} \\ \hline
Static 0$^{\circ}$ & 2.5 & 2.7 \\ \hline
Static 45$^{\circ}$ & 3.7 & 5.1 \\ \hline
Lidar & 8.8 & 10.1 \\ \hline
Gusts & 4.2 & 7.2 \\ \hline
\end{tabular}
\end{table}

\subsection{Wind Turbine}\label{sec:turbine}

\begin{figure}[t]
\centering
\begin{subfigure}[b]{0.48\textwidth}
\includegraphics[width=\textwidth]{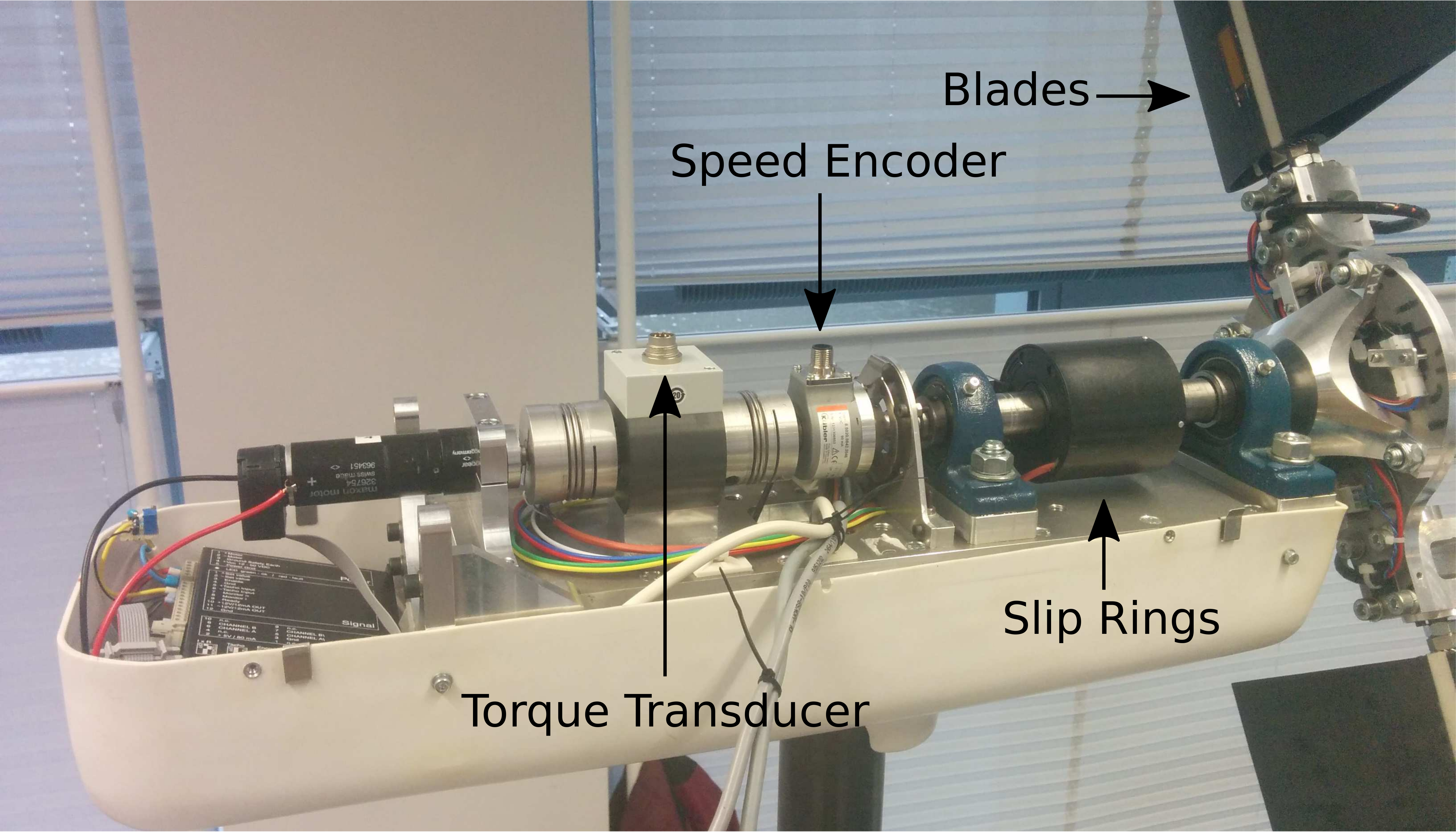}
\caption{Drive train}
\label{fig:drivetrain}
\end{subfigure}
\begin{subfigure}[b]{0.48\textwidth}
\includegraphics[width=\textwidth]{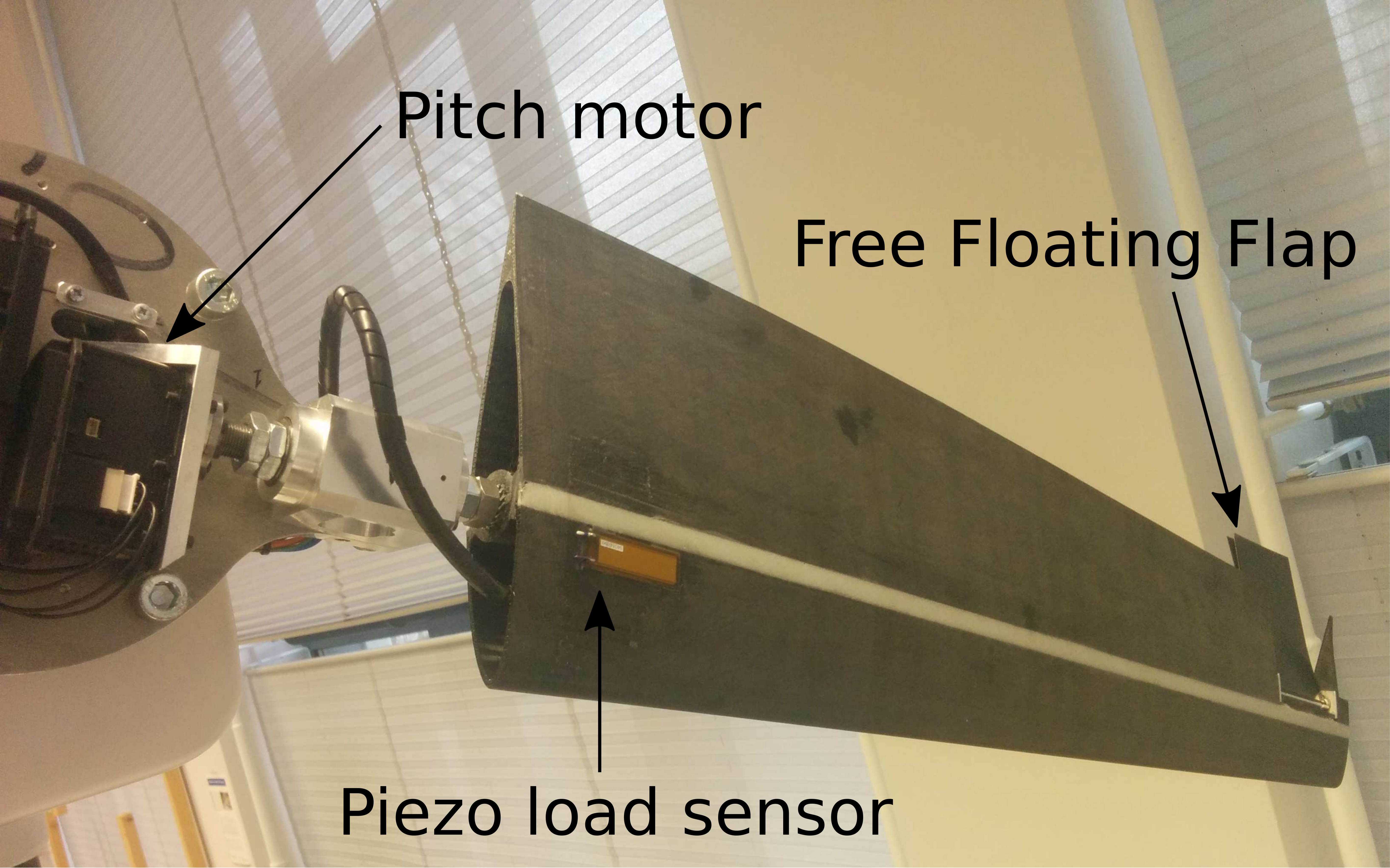}
\caption{Blade}
\label{fig:blade}
\end{subfigure}
\caption{Photographs of the drive train (a) and the blade (b) of the two-bladed wind turbine used for the experiments.}
\label{fig:pic}
\end{figure}

The wind turbine model that is used for these experiments is presented in \citep{SPRC2}. It is a two-bladed direct-drive wind turbine that is placed upwind of the wind tunnel. The drive train is shown in Figure \ref{fig:drivetrain}. The blades are connected with hub through a rigid connection with the shaft of Dynamixel MX-106 servomotors. These servomotors enable rotation of the blades around the longitudinal axis of the blade. The Dynamixel servomotors have a bandwidth of approximately 15$\,$Hz. The azimuth angle of the blades is measured through a position encoder located in the main shaft. For other experiments executed with this turbine, see \textit{e.g.} \cite{edwin}

The blades used for this experiments are designed and presented in \citep{flaps} and is shown in Figure \ref{fig:blade}. A Macro Fiber Composite (MFC) piezoelectric sensor is affixed to each blade, located at the root of the blade. These piezo's are used to measure the strain on the blades, which relates directly to the out-of-plane bending moments. 

With these blades and the wind conditions described in the previous subsection, rotor speeds of up to approximately 330$\,$rpm (5.5$\,$Hz) can be achieved. Considering the bandwidth of the servomotors, therefore periodic loads up to twice the rotor speed (2P) can be controlled.

Note that the blades also contain free-floating flaps that can be used for control by changing the input voltage of the MFC piezobenders attached to these flaps. These piezobenders have a much higher bandwidth than the servomotors, but the control authority is significantly lower. For results obtained with these flaps, see \textit{e.g.} \cite{flaps}. The experiments shown in this paper are obtained without using the piezobenders on the free-floating flaps as a control input. Furthermore, the wind turbine tower has free yaw capabilities, since it is mounted using two bearings. For the experiments performed for this paper, the yaw angle of the tower is fixed using a clamp.

To simulate the generator torque of the turbine, the generator is connected in series to a dump load (not shown in the figure). The generator torque is then controlled by setting the current to the dump load. 

\subsection{Real-time environment}\label{sec:rt}

As described above, the system contains 3 actuators (two servomotors controlling the pitch angle and the dump load controlling the generator torque) and sensors (measuring the loads on both blades and the azimuth angle). The communication between the sensors and actuators is realized through Simulink Real-Time \citep{slrt}. The desired controller is developed in MATLAB-Simulink \citep{matlab}, and subsequently compiled on a target computer. 

\begin{figure}[t]
\centering
\includegraphics[width=0.48\textwidth]{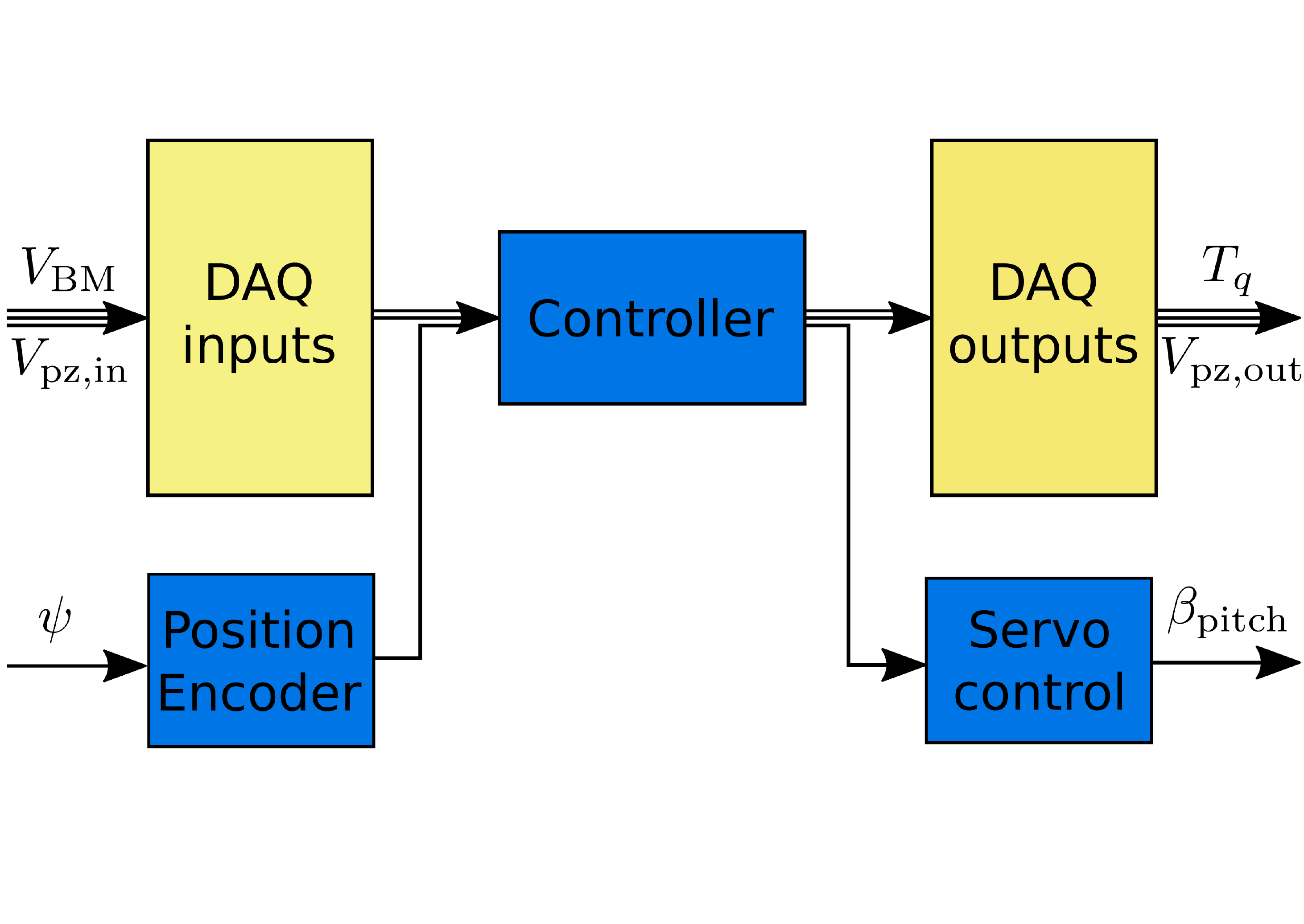}
\caption{A schematic representation of the interconnection between the data acquisition boards (DAQ's) and the controller. The blue blocks run at 200Hz, while the yellow blocks sample at 2kHz. The inputs are the voltages of the bending moments ($V_{\text{BM}}$) and the piezo flaps ($V_{\text{pz,in}}$), as well as the rotor azimuth $\psi$. Outputs are the generator torque $T_q$, the desired flap voltage $V_{\text{pz,out}}$ and the desired pitch angles $\beta_{\text{pitch}}$. Adapted from \cite{edwin}.}
\label{fig:DAQ}
\end{figure}

The target computer, an HP workstation Z600, communicates with the wind turbine through a National Instruments PCI-6259 data acquisition board (DAQ) as shown schematically in Figure~\ref{fig:DAQ}. The DAQ's have a sampling time of 2$\,$kHz, while the shaft position encoder and the Dynamixel servo motors operate at 200$\,$Hz. The controller is configured at the same sampling frequency, since the computation time of the SPRC algorithm on the target computer is slightly below 0.005$\,$s. With a more powerful target computer, it will most likely be possible to further decrease this computation time. To enable communication between the signals with different sampling frequencies, the Rate Transition functionality in Simulink is used, see \cite{slrt}. 

\section{Subspace Predictive Repetitive Control}\label{sec:sprc}
In this section, the Subspace Predictive Repetitive Control (SPRC) methodology is described. In the following section, the motivation for using SPRC is given. Subsequently, Section \ref{sec:ident} elaborates on the identification, and Section \ref{sec:rc} covers the Repetitive Control (RC) implementation.

\subsection{Motivation}\label{sec:why}

As mentioned in the introduction, the dominant frequencies of periodic wind turbine blade loads during operation are dependent on the rotor speed. The once-per-revolution load frequency is called 1P, and its higher harmonics 2P, 3P, etc. As these frequencies form the majority of the loads in wind turbines, the control effort can also be restricted to these frequencies. This can be achieved by using basis functions containing sinusoids of the required frequencies, as will be explained in Section \ref{sec:rc}.

Due to the periodic nature of the loads, RC is an effective methodology to handle these loads. RC determines the optimal control sequence for the next period, and iterates this process over time. Subsequently, the control signal will also adapt to changing operating conditions. This makes RC suitable for wind turbine control, as the wind flow is highly variable in real-world operating conditions. Varying rotor speed can be a problem for RC in turbine load control, since this essentially changes the period of the RC problem. However, in this paper, modifications to the RC algorithm will be presented that negate this problem.

To find the optimal RC sequence, a model of the system is necessary. By using data-driven subspace identification methods, the derived model is able to adapt to changing operating conditions. As new data becomes available, it will replace old data based on a forgetting factor $\lambda$. This procedure will be further explained in the next section. 

These combined features of SPRC make the methodology suitable for the task at hand. First, subspace identification will be done online, and the obtained system will be used to adapt the RC law. The optimal control sequence will then be implemented over the next rotation period to achieve the desired load disturbance rejection.

\subsection{Subspace Identification}\label{sec:ident}

The wind turbine system will be identified online using Markov parameters, and this identified system will then be used to establish a repetitive control law using basis functions. This method is similar to the one presented in \citep{SPRC1}, although essential additions have been made to improve performance for varying rotor speed. 

The wind turbine system is assumed to be represented by a discrete linear time-invariant (LTI) system with unknown periodic disturbances \citep{houtzager}\footnote{The framework is also able to work with periodic time-varying systems. For representation reasons, an LTI system is chosen here.}
\begin{align}
\label{eq:ssnormx}
x_{k+1} &= Ax_k + Bu_k +Ed_k +Ke_k \\
y_k &= Cx_k + Fd_k+e_k
\label{eq:ssnormy}
\end{align}
where $x_k\in \mathbb{R}^n$ is the state vector, $u_k\in\mathbb{R}^r$ the input vector; in this case the pitch angles of both blades. The output vector $y_k\in\mathbb{R}^l$ contains the blade loads as measured by the MFC's mounted on the blades. Disturbance $d_k\in\mathbb{R}^m$ represents the periodic component on the load of the blades, and $e_k\in\mathbb{R}^l$ the aperiodic component. Rewriting these equations in descriptor form, the following state-space equations are obtained
\begin{align}
\label{eq:ssinox}
x_{k+1}&=\tilde{A}x_k+Bu_k+\tilde{E}d_k+Ky_k\\
y_k&=Cx_k+Fd_k+e_k
\label{eq:ssinoy}
\end{align}
with $\tilde{A} = A-KC$ and $\tilde{E}=E-KF$. By defining the difference operator $\delta$, we obtain:
\begin{equation*}
\delta d_k = d_k - d_{k-P} = 0
\end{equation*}
Similarly, $\delta u$, $\delta y$ and $\delta e$ can be defined. Applying the $\delta$-notation on the innovation system yields a representation where the $d$-term disappears:
\begin{align}
\label{eq:ssdiffx}
\delta x_{k+1} &= \tilde{A}\delta x_k + B\delta u_k +K\delta y_k \\
\delta y_k &= C\delta x_k +\delta e_k
\label{eq:ssdiffy}
\end{align}
Next, the stacked vector $\delta U_k^{(p)}$ for a given past window $p$ is defined as:

\begin{equation}
\delta U_k^{(p)} = \begin{bmatrix}
u_k - u_{k-P} \\ u_{k+1} - u_{k+1-P} \\ \vdots \\ u_{k+p-1} - u_{k+p-1-P}
\end{bmatrix}
\end{equation}
and similarly $\delta Y_k^{(p)}$. Then, by elevating \eqref{eq:ssdiffx}, the state vector $\delta x_{k+p}$ can be written as:
\begin{equation*}
\delta x_{k+p} = \tilde{A}^p\delta x_k + \begin{bmatrix}
\mathscr{K}_u^{(p)} & \mathscr{K}_y^{(p)}
\end{bmatrix}\begin{bmatrix}
\delta U_k^{(p)} \\ \delta Y_k^{(p)}
\end{bmatrix}
\end{equation*}
with:
\begin{align}
\mathscr{K}_u^{(p)} &= \begin{bmatrix}
\tilde{A}^{p-1}B & \tilde{A}^{p-2}B & \dots & B 
\end{bmatrix}\\
\mathscr{K}_y^{(p)} &= \begin{bmatrix}
\tilde{A}^{p-1}{K} & \tilde{A}^{p-2}{K} & \dots & {K} 
\end{bmatrix}
\end{align}
Here, similar to \cite{houtzager}, it is assumed that the system given in Equations~\eqref{eq:ssdiffx}~and~\eqref{eq:ssdiffy} is assymptotically stable, controllable and observable. It is important to select $p$ sufficiently large, such that $\tilde{A}^j \approx 0 \forall j\geq p$, \citep{chiuso}. For such  $p$, the equation above can be simplified to:

\begin{equation}
\delta x_{k+p} \approx \begin{bmatrix}
\mathscr{K}_u^{(p)} & \mathscr{K}_y^{(p)}
\end{bmatrix}\begin{bmatrix}
\delta U_k^{(p)} \\ \delta Y_k^{(p)}
\end{bmatrix}
\label{eq:dxdef}
\end{equation}
Substituting this result into \eqref{eq:ssdiffy} yields

\begin{equation}
\delta y_k \approx \begin{bmatrix}
C\mathscr{K}_u^{(p)} & C\mathscr{K}_y^{(p)}
\end{bmatrix}\begin{bmatrix}
\delta U_k^{(p)} \\ \delta Y_k^{(p)}
\end{bmatrix}+\delta e_k
\label{eq:dydeff}
\end{equation}
During online identification, the values of the parameters $C\mathscr{K}$ are estimated based on the measurements $y$ and $u$. These parameters define the behavior of the wind turbine system, and are called the Markov parameters $\Xi \in \mathbb{R}^{l\times((r+l)\cdot p))}$

\begin{equation}
\Xi= \begin{bmatrix}
C\mathscr{K}_u^{(p)} & C\mathscr{K}_y^{(p)}
\end{bmatrix}
\label{eq:markovdef}
\end{equation}
A batchwise computation of the Markov estimates $\hat{\Xi}$ at time instant $k$ is then performed by finding the unique solution to the least-squares equation

\begin{equation}
\hat{\Xi}_k = \arg \min_{\hat{\Xi}_k} \sum^k_{i=-\infty} \begin{Vmatrix}
\delta y_i - \lambda\hat{\Xi}_k\begin{bmatrix}
\delta U_{i-p}^{(p)} \\ \delta Y_{i-p}^{(p)}
\end{bmatrix}
\end{Vmatrix}_2^2
\label{eq:markovest}
\end{equation}
In this algorithm, a forgetting factor $\lambda$ of between 0 and 1 is introduced to adapt to changes in the system dynamics. To improve the robustness of the identification, a large value (\textit{e.g.} $\lambda = 0.99999$) is chosen, which, as a rule of thumb, represents a window of $10^6$ samples \citep{gustafsson}. Subsequently, the summation given in \eqref{eq:markovest} no longer needs an infinite past window. From the definition of $\Xi$ as shown in \eqref{eq:markovdef}, it follows that $\hat{\Xi}$ at time instant $k$ contains estimates of the following matrices:

\begin{equation}
\begin{aligned}
\hat{\Xi}_k =&\left[ \begin{smallmatrix}
\widehat{CA^{p-1}B} & \widehat{CA^{p-2}B} & \dots & \widehat{CB}&\widehat{CA^{p-1}K} & \widehat{CA^{p-2}K} & \dots & \widehat{CK}
\end{smallmatrix}\right]_k
\end{aligned}
\end{equation}
It is important that the input of the system is persistently exciting and of a sufficiently high order, in order to guarantee a unique solution of the least-squares problem \eqref{eq:markovest} \citep{verhaegen}. The recursive equivalent of this problem is then solved using a QR recursive least-squares algorithm as presented in \citep{vdveen}. 

Typically, adaptive control methodologies that combine online identification with simultaneous control can not guarantee certain stability and performance characteristics \citep{stability}. Therefore, the method proposed in \cite{SPRC2} to first run the controller in identification phase at the beginning of each experiment is used.

\subsection{Repetitive Control}\label{sec:rc}

For repetitive control, the output needs to be predicted over period $P$, with $P \geq p$ but usually $P \gg p$. To achieve this, the output equation needs to be lifted over $P$ to obtain $\delta P^{(P)}_{k+P}$. For this purpose, the Toeplitz matrix $\tilde{H}^{(P)}\in \mathbb{R}^{(l\cdot P)\times(l\cdot P)}$ and the extended observability matrix $\tilde{\Gamma}^{(P)} \in \mathbb{R}^{(l\cdot P)\times n}$ are defined:

\begin{equation}
\tilde{H}^{(P)} = \begin{bmatrix}
0 & 0 & 0 & \dots \\
CB & 0 & 0 & \dots \\
C\tilde{A}B & CB & 0 & \dots \\
\vdots & \vdots & \ddots & \vdots \\
C\tilde{A}^{p-1}B & C\tilde{A}^{p-2}B & C\tilde{A}^{p-3}B & \dots \\
0 & C\tilde{A}^{p-1}B & C\tilde{A}^{p-2}B & \dots \\
0 & 0 & C\tilde{A}^{p-1}B & \ddots \\
\vdots & \vdots & \ddots & \ddots \\
\end{bmatrix}
\label{eq:Htdef}
\end{equation}

\begin{equation}
\tilde{\Gamma}^{(P)} = \begin{bmatrix}
C \\ C\tilde{A} \\ C\tilde{A}^2 \\ \vdots \\ C\tilde{A}^p \\ 0 \\ \vdots \\ 0
\end{bmatrix}
\label{eq:Gamtdef}
\end{equation}
Similarly, $H^{(P)}$ and $\Gamma^{(P)}$ are defined by replacing all $\tilde{A}$ by $A$. Likewise, $\tilde{G}^{(P)}$ is defined by replacing $B$ by $K$ in $\tilde{H}^{(P)}$. Using these matrices, the lifted output equation can be written as

\begin{equation}
\delta Y^{(P)}_{k+P} = \tilde{\Gamma}^{(P)} \delta x_{k+P} + \begin{bmatrix}
\tilde{H}^{(P)} & \tilde{G}^{(P)}
\end{bmatrix}\begin{bmatrix}
\delta U^{(P)}_{k+P} \\ \delta Y^{(P)}_{k+P}
\end{bmatrix}
\end{equation}
Substituting the approximation of $\delta x_k$ as given in \eqref{eq:dxdef} yields

\begin{equation}
\begin{aligned}
\delta Y^{(P)}_{k+P} = \tilde{\Gamma}^{(P)} &\begin{bmatrix}
\mathscr{K}_u^{(P)} & \mathscr{K}_y^{(P)}
\end{bmatrix}\begin{bmatrix}
\delta U_k^{(P)} \\ \delta Y_k^{(P)}
\end{bmatrix}\\
& + \begin{bmatrix}
\tilde{H}^{(P)} & \tilde{G}^{(P)}
\end{bmatrix}\begin{bmatrix}
\delta U^{(P)}_{k+P} \\ \delta Y^{(P)}_{k+P}\end{bmatrix}
\end{aligned}
\label{eq:dYdef}
\end{equation}
Notice that the first $(P-p)\cdot r$ columns of $\mathscr{K}_u^{(P)}$ and $\mathscr{K}_y^{(P)}$ are $0$. It is also key to note that all the matrices from \eqref{eq:dYdef} can be constructed by using the elements of the Markov estimates $\hat{\Xi}$. The future output $Y^{(P)}_{k+P}$ are then predicted using the previous outputs $Y^{(P)}_k$ and previous and future inputs $U_k^{(P)}$ and $U_{k+P}^{(P)}$. Subsequently, \eqref{eq:dYdef} can be rewritten as:

\begin{equation}
\begin{split}
Y^{(P)}_{k+P} = \begin{bmatrix} I_{l\cdot P} &
\widehat{\Gamma^{(P)} \mathscr{K}^{(P)}_u} & \widehat{\Gamma^{(P)} \mathscr{K}^{(P)}_y}
\end{bmatrix}\begin{bmatrix}
Y^{(P)}_k \\ \delta U_k^{(P)} \\ \delta Y^{(P)}_k
\end{bmatrix}\\ + \hat{H}^{(P)}\delta U^{(P)}_{k+P}
\end{split}
\label{eq:Yhatdef}
\end{equation}
This result is obtained by using the following equalities:

\begin{equation*}
\begin{aligned}
\left(I-\tilde{G}^{(P)}\right)^{-1}\tilde{\Gamma}^{(P)} & = \Gamma^{(P)}\\
\left(I-\tilde{G}^{(P)}\right)^{-1}\tilde{H}^{(P)} &= H^{(P)}
\end{aligned}
\end{equation*}
Subsequently, the system is transformed into a state space representation, such that classic state feedback control can be applied \citep{hallouzi}:

\begin{equation}
\begin{split}
\underbrace{\left[\begin{smallmatrix}
Y^{(P)}_{k+P} \\ \delta U_{k+P}^{(P)} \\ \delta Y^{(P)}_{k+P}
\end{smallmatrix}\right]}_{\hat{\mathscr{X}}_{k+P}} = 
\underbrace{\left[\begin{smallmatrix}
I_{l\cdot P} & \widehat{\Gamma^{(P)}\mathscr{K}^{(P)}_u} & \widehat{\Gamma^{(P)}\mathscr{K}^{(P)}_y} \\
0_{(r\cdot P)\times(l\cdot P)} & 0_{r\cdot P} & 0_{(r\cdot P)\times(l\cdot P)} \\
0_{l\cdot P} & \widehat{\Gamma^{(P)}\mathscr{K}^{(P)}_u} & \widehat{\Gamma^{(P)}\mathscr{K}^{(P)}_y}
\end{smallmatrix}\right]}_{\hat{\mathscr{A}}_k}\underbrace{\left[\begin{smallmatrix}
Y^{(P)}_{k} \\ \delta U_{k}^{(P)} \\ \delta Y^{(P)}_{k}
\end{smallmatrix}\right]}_{\hat{\mathscr{X}}_{k}} \\
 + \underbrace{\left[\begin{smallmatrix}
\hat{H}^{(P)} \\ I_{r\cdot P} \\ \hat{H}^{(P)}
\end{smallmatrix}\right]}_{\hat{\mathscr{B}}_k}\begin{smallmatrix}\delta U^{(P)}_{k+P}\end{smallmatrix}
\end{split}
\label{eq:sshat}
\end{equation}
Next, a state feedback controller is synthesized using a discrete algebraic Riccati equation (DARE). As mentioned in Section \ref{sec:why}, the goal of this IPC implementation is to target the 1P and 2P loads. As a result, the control signal should only contain these frequencies. To achieve this, a basis function projection is proposed such that $U_k$ only contains sinusoids of the desired frequencies. This is accomplished by using the following transformation matrix $\phi \in \mathbb{R}^{(r\cdot P)\times (4r)}$:

\begin{equation}
\phi = \begin{bmatrix}
\sin\frac{2\pi}{P} & \cos\frac{2\pi}{P} & \sin\frac{4\pi}{P} & \cos\frac{4\pi}{P}\\
\sin\frac{4\pi}{P} & \cos\frac{4\pi}{P} & \sin\frac{8\pi}{P} & \cos\frac{8\pi}{P}\\
\vdots & \vdots & \vdots & \vdots \\
\sin 2\pi & \cos 2\pi & \sin 4\pi & \cos 4\pi
\end{bmatrix}\otimes I_r
\label{eq:phidef}
\end{equation}
where the symbol $\otimes$ represent the Kronecker product. Considering that the bandwidth of the pitch motors limits the control authority to the 1P and 2P frequencies, only these frequencies will be present. Notice that by taking a linear combination of the sinusoids in this matrix, a control signal containing only the desired 1P and 2P frequencies is obtained. The control input $U_k$ is determined using:

\begin{equation}
U_k^{(P)} = \phi\theta_j
\label{eq:thetadef}
\end{equation}
where the subscript $j$ represents the rotation count. Subsequently, the vector $\theta\in \mathbb{R}^{4r}$, that determines the amplitude and phase of the sinusoids, is updated every rotation period $P$. 

Note that the system of \eqref{eq:sshat} is quite high-dimensional, as $\hat{\mathscr{A}} \in \mathbb{R}^{\left((2l+r)\cdot P\right)\times \left((2l+r)\cdot P\right)}$. Apart from limiting the frequency content of the control signal, the transformation also reduces the dimensionality of the DARE, as $\theta$ only contains $4r$ elements, substantially reducing the computational load of the problem.

As the pitch angles are now limited to sinudoidal signals with frequencies 1P and 2P, and the system is assumed to be linear over one period $P$, the load signals $Y_k$ will also be limited to these frequencies. As a result, we can transform this signal using the same transformation matrix:

\begin{equation}
Y_k = \phi \bar{Y}_j
\label{eq:Ybardef}
\end{equation}
Note that for this transformation to be possible, the number of inputs needs to be equal to the number of outputs, \textit{i.e.} $r = l$. However, this does not limit the possibilities of the algorithm for load alleviation, since generally the outputs are chosen as the root bending moments of each blade, and the inputs are the pitch angles of the blades. Consequently, both $r$ and $l$ equal the number of blades.

Similar to \eqref{eq:Ybardef}, the lower dimensional signals can be found by using the inverse transformation $\phi^+$, where $+$ represents the Moore-Penrose pseudoinverse:

\begin{equation}
\theta_j = \phi^+ U_k, \quad \bar{Y}_j = \phi^+ Y_k
\label{eq:projdef}
\end{equation}
Using \eqref{eq:thetadef} and \eqref{eq:Ybardef}, we can rewrite \eqref{eq:sshat} in the following lower dimensional form:

\begin{equation}
\begin{split}
\underbrace{\left[\begin{smallmatrix}
\bar{Y}^{(P)}_{j+1} \\ \delta \theta_{j+1}^{(P)} \\ \delta \bar{Y}^{(P)}_{j+1}
\end{smallmatrix}\right]}_{\hat{\bar{\mathscr{X}}}_{j+1}} = 
\underbrace{\left[\begin{smallmatrix}
I_{l\cdot P} & \phi^+\widehat{\Gamma^{(P)}\mathscr{K}^{(P)}_u}\phi & \phi^+\widehat{\Gamma^{(P)}\mathscr{K}^{(P)}_y}\phi \\
0_{l\cdot P} & 0_{r\cdot P} & 0_{l\cdot P} \\
0_{l\cdot P} & \phi^+\widehat{\Gamma^{(P)}\mathscr{K}^{(P)}_u}\phi & \phi^+\widehat{\Gamma^{(P)}\mathscr{K}^{(P)}_y}\phi
\end{smallmatrix}\right]}_{\hat{\bar{\mathscr{A}}}_j}\underbrace{\left[\begin{smallmatrix}
\bar{Y}^{(P)}_{j} \\ \delta \theta_{j}^{(P)} \\ \delta \bar{Y}^{(P)}_{j}
\end{smallmatrix}\right]}_{\hat{\bar{\mathscr{X}}}_{j}} \\
 + \underbrace{\left[\begin{smallmatrix}
\phi^+\hat{H}^{(P)}\phi \\ I_{r\cdot P} \\ \phi^+\hat{H}^{(P)}\phi
\end{smallmatrix}\right]}_{\hat{\bar{\mathscr{B}}}_j}\begin{smallmatrix}\delta \theta^{(P)}_{j+1}\end{smallmatrix}
\end{split}
\label{eq:ssproj}
\end{equation}
The size of this projected matrix $\bar{\mathscr{A}} \in \mathbb{R}^{12l\times 12l}$ is significantly smaller than the original matrix $\mathscr{A} \in \mathbb{R}^{3lP\times3lP}$. As usually $P\gg 4$, using a basis function transformation significantly reduces the order of the optimization problem. Moreover, the transformation guarantees that the input $U_k$ is a smooth signal with the desired frequency content.

Next, a state feedback control problem is solved to determine the control input $\theta$. The state feedback gain is obtained by minimizing the following quadratic cost function

\begin{equation}
\displaystyle J = \sum^{\infty}_{j=0} \left|\left| \left(\bar{\mathscr{X}}_{j}\right)^T Q \bar{\mathscr{X}}_{j} + \left(\delta\theta_j\right)^T R \delta\theta_j \right|\right|^2_2
\label{eq:lqrcost}
\end{equation}
where $Q$ and $R$ are weighing matrices for the state and input vector respectively. As in LQR problems, the state feedback control gain can be found by solving the DARE at iteration $j$ using an initial estimate of $P_R$:

\begin{equation*}
\begin{aligned}
P_{R,j+1} = Q + &\bar{\mathscr{A}}^T_j(P_{R,j}- P_{R,j}\bar{\mathscr{B}}^T_j\\
& \times (R+\bar{\mathscr{B}}^T_jP_{R,j}\bar{\mathscr{B}}_j)^{-1} \bar{\mathscr{B}}^T_jP_{R,j})\bar{\mathscr{A}}_j
\end{aligned}
\end{equation*}
Subsequently, the optimal state feedback gain $K_f$ is defined as:

\begin{figure*}[t]
\centering
\includegraphics[width=\textwidth]{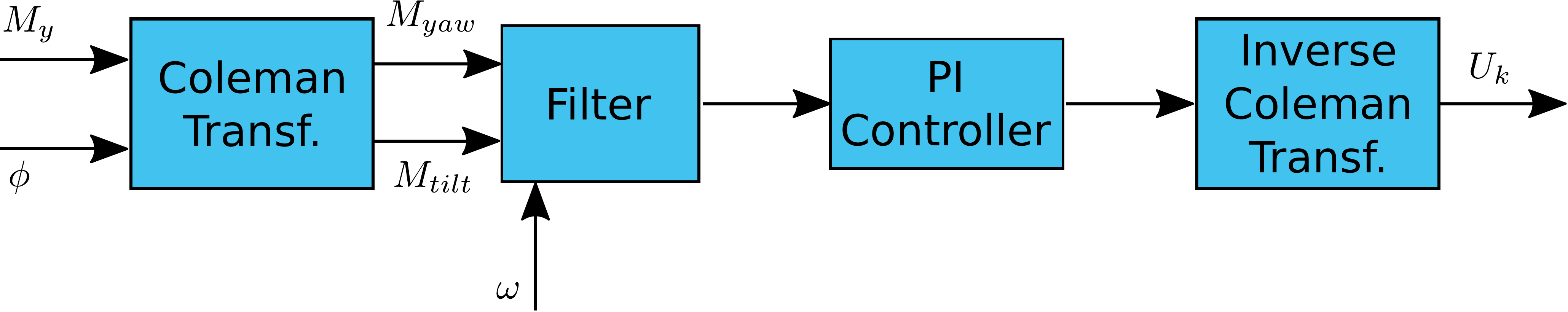}
\caption{A schematic representation of the conventional IPC algorithm \citep{ipc1}, that will be used as a benchmark individual pitch controller in this paper.}
\label{fig:CIPC}
\end{figure*}

\begin{equation*}
K_{f,j} = \left(R+\bar{\mathscr{B}}^T_jP_{R,j}\bar{\mathscr{B}}_j\right)^{-1}\bar{\mathscr{B}}^T_jP_{R,j}\bar{\mathscr{A}}
\end{equation*}
Now, it is possible to determine the control input vector $\delta \theta_j$, which, after a transformation, can be implemented on the wind turbine. Using the state feedback law:

\begin{equation}
\delta \theta_{j+1} = -K_{f,j}\bar{\mathscr{X}}_j
\end{equation}
Then, using $\delta \theta_{j+1} = \theta_{j+1}-\theta_j$ and introducing variables $\alpha$ and $\beta$:

\begin{equation}
\theta_{j+1} = \alpha\theta_j-\beta K_{f,j} \begin{bmatrix}
\bar{Y}_j \\ \delta\theta_j \\ \delta\bar{Y}_j
\end{bmatrix}
\end{equation}
In order to add the possibility to manipulate the convergence characteristics of the algorithm, the tuning parameters $\alpha$ and $\beta$ are included. Both $\alpha$ and $\beta$ are defined in the interval $[0,1]$, and give a weight on new and older data respectively. The input signal $U_k$ can now be determined by using the inverse basis function transformation as given in \eqref{eq:thetadef}. 

One problem that needs to be dealt with, is the potential variation of rotor speed due to, \textit{e.g.}, wind turbulence or changing inflow wind speed. Therefore, a phase shift between input and output could occur. To prevent this, the rotor azimuth $\psi_k$ measured through the shaft encoder, which is equal to the angle ${2\pi k}/{P}$ at time instant $k$, can be used. As a result, the algorithm is also able to account for variations in rotor velocity. In this paper, the parameter $P$, which represents one full rotation, is chosen slightly smaller than the expected rotation period, in order to guarantee a new control sequence at the end of each rotation. This sequence is then implemented when the rotation is completed. The control input at time instant $k$ now becomes
\begin{equation}
u_k = \left(\begin{bmatrix}
\sin \psi_k &\cos \psi_k &\sin 2\psi_k &\cos 2\psi_k
\end{bmatrix}\otimes I_r\right)\theta_j
\label{eq:ukdef}
\end{equation}
where input $u_k \in \mathbb{R}^{r}$ represents the individual pitch angles that are implemented on the wind turbine system at time instant $k$.

\subsection{Benchmark controller}\label{sec:cipc}

As a benchmark load alleviation controller, Conventional Individual Pitch Control (CIPC) will be used, first introduced in \cite{ipc1}. In this approach, the Coleman transformations \citep{coleman} are used to obtain the yaw and tilt moments on the rotor plane. Subsequently, a notch filter and PI-controller are applied, followed by the inverse Coleman transformation, to determine the individual pitch actions. Figure~\ref{fig:CIPC} shows a schematic representation of this control methodology. 

In terms of controller implementation, CIPC can be considered less complex than SPRC. The main reason for this is the fact that no system identification is necessary, instead using the relatively straightforward Coleman transformations. However, similar to the SPRC algorithm, some controller parameters do need to be tuned in CIPC to guarantee performance. In this case, these parameters are the gains of the PI-controllers shown in Figure~\ref{fig:CIPC}. 

\section{Results}\label{sec:results}

In this section, the control methodology presented in Section \ref{sec:sprc} will be evaluated in the wind tunnel setup presented in Section \ref{sec:setup}. As discussed in Section \ref{sec:grid}, four different wind conditions have been studied for three different wind speeds. First, the results for constant wind conditions will be presented, followed by the experiments with changing wind conditions.

\subsection{Constant operating conditions}\label{sec:constant}

In this section, the results of the SPRC IPC implementation on the scaled wind turbine in constant operating conditions are presented. All figures shown are for an inflow wind speed of 5$\,$m/s.

SPRC will be compared with Conventional IPC \citep{ipc1} to evaluate the performance of the control algorithm. This is done by executing 120$\,$s experiments for both control strategies, as well as a baseline experiment with no IPC. For clarity, the time domain figures show the loads over smaller time intervals, whereas the power spectra and load reductions are determined using the data of the entire 120$\,$s interval. With a sampling interval of 200$\,$Hz, this results in data sets of 24000 load measurements.

\begin{figure}[t]
\includegraphics[width=0.49\textwidth]{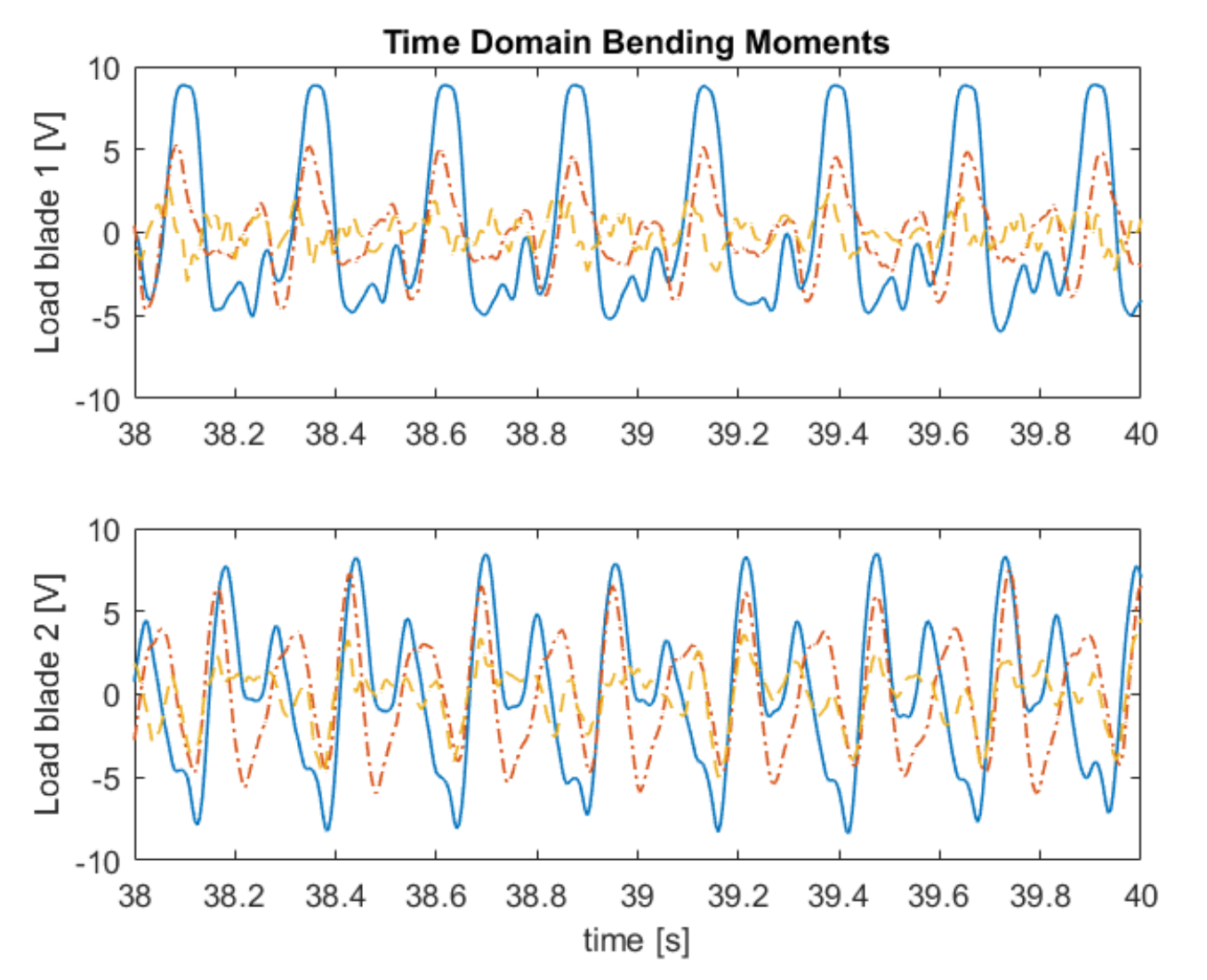}
\caption{The variations of the loads on both blades of the turbine for three different situations: no control ({\color{cyan}------}), CIPC ({\color{red}-$\cdot$-$\cdot$-$\cdot$-}) and SPRC ({\color{Dandelion}- - - -}). The inflow wind speed is 5$\,$m/s with a static 0$^{\circ}$ grid configuration (centerline TI: 2.5$\%$), resulting in a rotor speed of 230$\,$rpm.}
\label{fig:tdopen}
\end{figure}

The loads on the blades for the three previously introduced strategies, for the static 0$^{\circ}$ grid mode, are shown in Figure \ref{fig:tdopen}. This figure shows that both methods significantly decrease the periodic loads. With CIPC, the variance of the blade loads is reduced with 61.7$\%$, while with SPRC the reduction is even larger: 86.8$\%$.

\begin{figure}[t]
\includegraphics[width=0.49\textwidth]{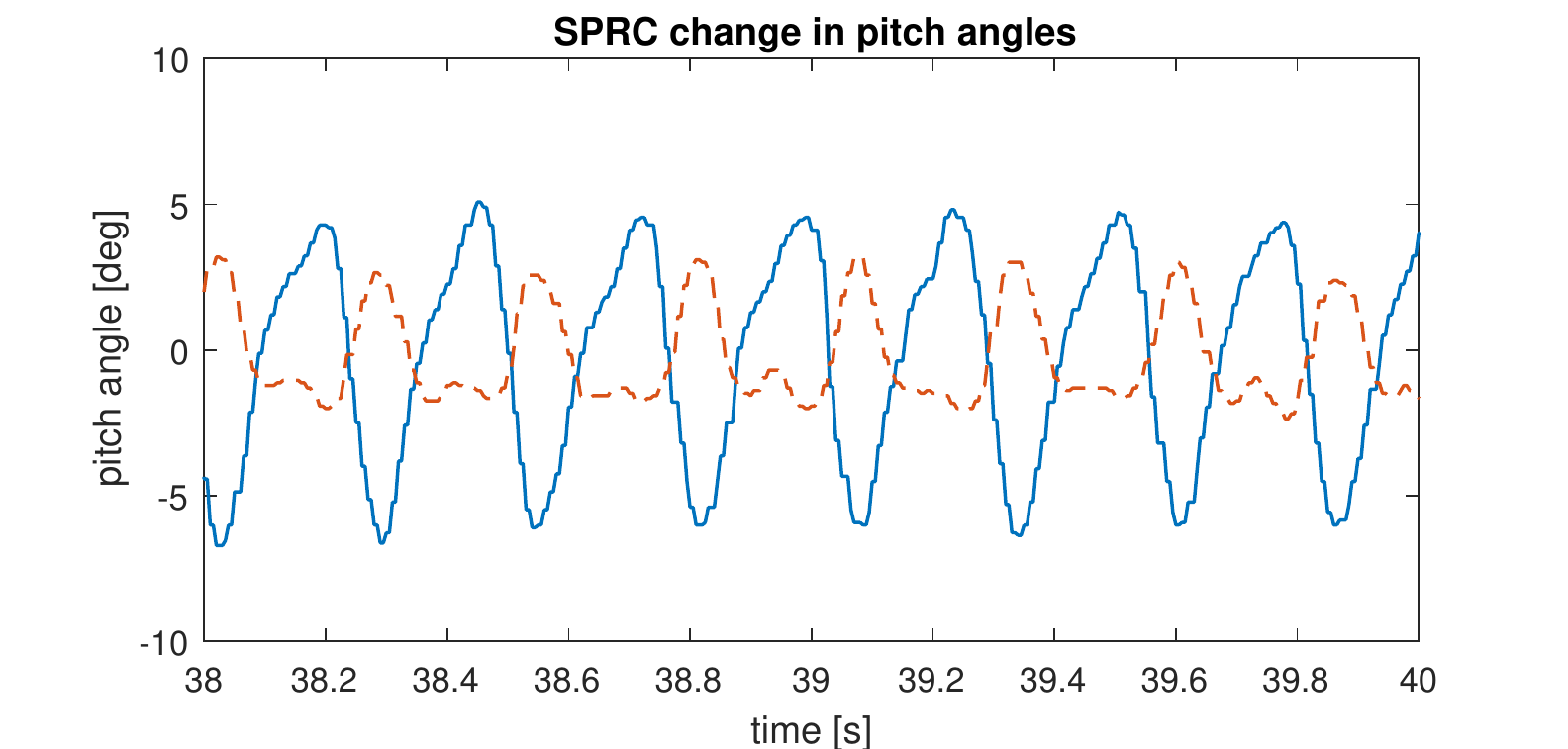}
\caption{The individual pitch angles for blade 1 ({\color{cyan}------}) and blade 2 ({\color{red}- - - -}) as applied on the turbine by the SPRC algorithm. The inflow wind speed is 5$\,$m/s with a static 0$^{\circ}$ grid configuration (centerline TI: 2.5$\%$), resulting in an rotor speed of 230$\,$rpm.}
\label{fig:tdpitch}
\end{figure}

Figure \ref{fig:tdpitch} shows the individual pitch action of the blades when SPRC is applied. Clearly, the signals are not symmetrical. This is caused by the rotor imbalance and other system imperfections, which SPRC accounts for by generating pitch angles for each blade individually. These signals are constructed using exclusively sinusoids of 1P and 2P frequency.

\begin{figure}[t]
\includegraphics[width=0.49\textwidth]{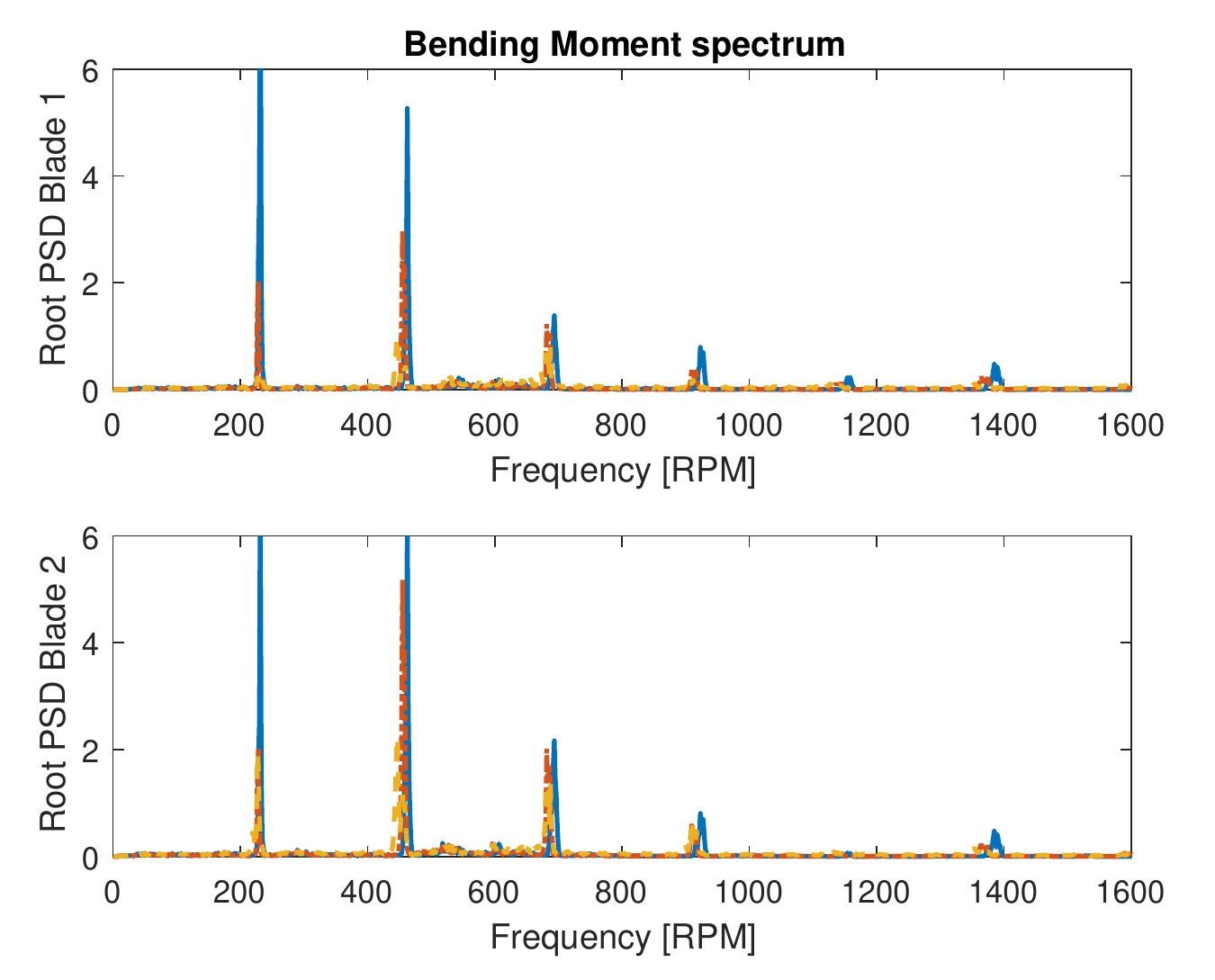}
\caption{The power spectrum of the blade loads showing the periodic loads for three different situations: no control ({\color{cyan}------}), CIPC ({\color{red}-$\cdot$-$\cdot$-$\cdot$-}) and SPRC ({\color{Dandelion}- - - -}). The inflow wind speed is 5$\,$m/s with a static 0$^{\circ}$ grid configuration (centerline TI: 2.5$\%$), resulting in an rotor speed of 230$\,$rpm.}
\label{fig:fdopen}
\end{figure}

The frequency domain plot of the signals shown in Figure \ref{fig:tdopen} are depicted in Figure \ref{fig:fdopen}. As expected, this figure shows large peaks at the frequencies 1P and 2P. At higher harmonics (3P, 4P, etc.) these peaks become significantly smaller, validating the choice to only apply control on the 1P and 2P frequencies. Figure \ref{fig:fdopen} also shows that SPRC achieves a substantial reduction of the 1P and 2P loads compared to both the baseline case and Conventional IPC.

It is clear that at constant, low turbulent conditions, SPRC achieves a larger load reduction than conventional IPC. These results are in line with the experiments with no turbulence done by \citep{SPRC2}. In the following, it will be shown that positive results can also be achieved at higher turbulent wind conditions generated by the active grid.

\begin{figure}[t]
\includegraphics[width=0.49\textwidth]{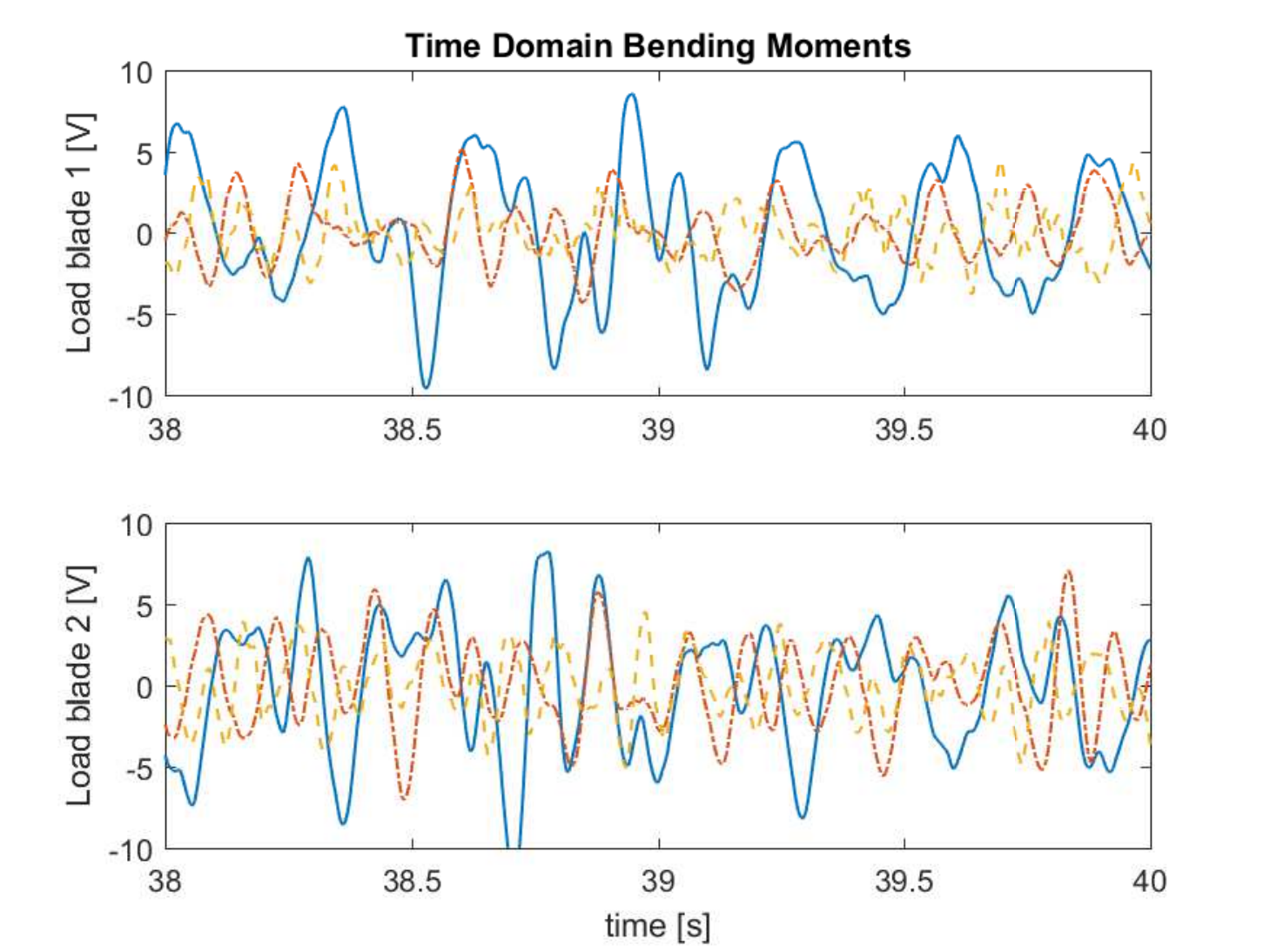}
\caption{The variations of the loads on both blades of the turbine for three different situations: no control ({\color{cyan}------}), CIPC ({\color{red}-$\cdot$-$\cdot$-$\cdot$-}) and SPRC ({\color{Dandelion}- - - -}). The inflow wind speed is 5$\,$m/s with the lidar grid mode (centerline TI: 8.8$\%$), resulting in an rotor speed of 210$\,$rpm.}
\label{fig:tdlidar}
\end{figure}

The excitation protocol that generates the highest turbulence is the lidar mode (see Table \ref{tab:modes}). As a result, significantly higher blade loads are expected for this mode compared to the results shown above. This can also be observed in Figure \ref{fig:tdlidar}. The peak loads in this figure are more irregular than in Figure \ref{fig:tdopen} due to the loads induced by turbulence. Nontheless, both control strategies still clearly produce load reductions, although it is less clear to see which of the two performs better. Evaluating the variance of the blade loads shows a reduction of 57.0$\%$ for CIPC and 65.1$\%$ for SPRC. 

\begin{figure}[t]
\includegraphics[width=0.49\textwidth]{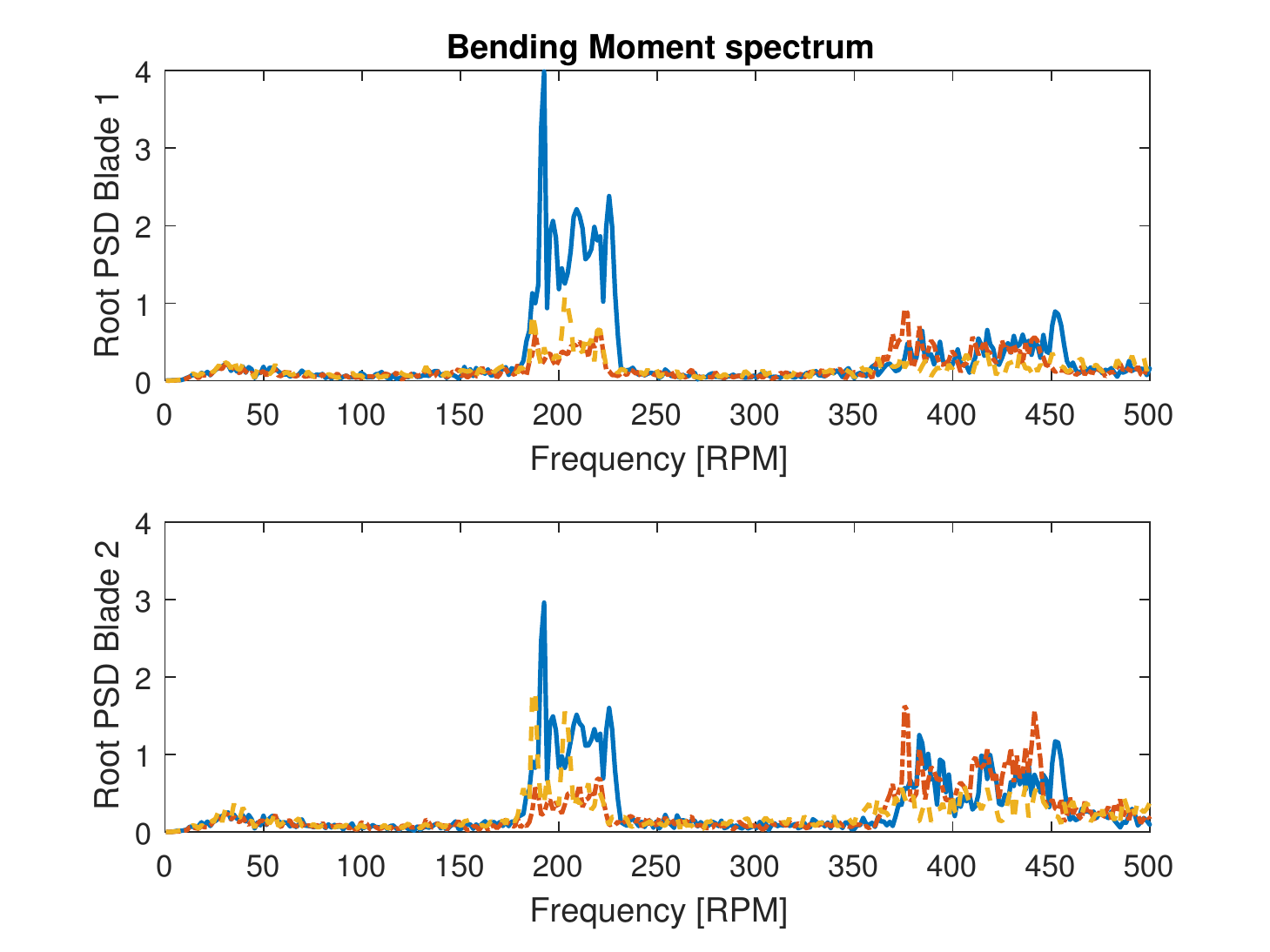}
\caption{The power spectrum of the blade loads showing the 1P and 2P loads for three different situations: no control ({\color{cyan}------}), CIPC ({\color{red}-$\cdot$-$\cdot$-$\cdot$-}) and SPRC ({\color{Dandelion}-$\cdot$-$\cdot$-$\cdot$-}). The inflow wind speed is 5$\,$m/s with the lidar grid mode (centerline TI: 8.8$\%$), resulting in an rotor speed of approximately 210$\,$rpm.}
\label{fig:fdlidar}
\end{figure}

The power spectrum of these measurements with the active grid in lidar mode are shown in Figure \ref{fig:fdlidar}. Notice that the rotor speed slightly decreased compared to Figure \ref{fig:fdopen}, from 230$\,$rpm to 210$\,$rpm. This can be explained by the active grid: the inflow velocity is 5$\,$m/s \textit{before} the active grid. As the grid is enabled, it reduces the wind velocity perpendicular to the turbine, resulting in a decrease of the rotor speed. 

Figure \ref{fig:fdlidar} also exhibits a much broader peak around the 1P and 2P frequencies than the low-TI case. Due to the turbulence, the rotor speed fluctuates more. Consequently, the power spectrum shows a 1P peaks over a broad range of frequencies in which the rotor speed moves. Despite the changing rotor speed, Figure \ref{fig:fdlidar} clearly shows a reduction of the 1P loads attained by both controllers, and SPRC also achieves a 2P load reduction.

\begin{table}[b]
\centering
\begin{minipage}{\textwidth}
\caption{Load reductions compared to baseline (no control) for all investigated inflow conditions. The numbers indicate the reduction of the variance of the load in $\%$ for all 4 grid modes. The different inflow velocities of the experiments (4, 4.5 and 5$\,$m/s) are also given.}
\label{tab:results}
\begin{tabular}{lcccccccccccc}
\hline
 & \multicolumn{3}{c}{\textbf{Static 0$^{\circ}$} (TI: 2.5$\%$)} & \multicolumn{3}{c}{\textbf{Static 45$^{\circ}$} (TI: 3.7$\%$)} & \multicolumn{3}{c}{\textbf{Lidar} (TI: 8.8$\%$)} & \multicolumn{3}{c}{\textbf{Gusts} (TI: 4.2$\%$)} \\
\cline{2-13}
 & \textbf{4$\,$m/s} & \textbf{4.5$\,$m/s} & \textbf{5$\,$m/s} & \textbf{4$\,$m/s} & \textbf{4.5$\,$m/s} & \textbf{5$\,$m/s} & \textbf{4$\,$m/s} & \textbf{4.5$\,$m/s} & \textbf{5$\,$m/s} & \textbf{4$\,$m/s} & \textbf{4.5$\,$m/s} & \textbf{5$\,$m/s}  \\
\hline
\textbf{CIPC [$\%$]} & 38.7 & 56.8 & 61.7 & 55.9 & 74.4 & 37.1 & 17.6 & 50.6 & 57.0& 38.0 & 52.8 & 47.7 \\
\hline
\textbf{SPRC} & \multicolumn{12}{c}{} \\ \hline
\textbf{1P [$\%$]} & 57.3 & 53.1 & 61.7 & 1.1 & 62.6 & 47.2 & 17.4 & 44.4 & 23.4 & -6.4 & 84.4 & 72.0 \\ \hline 
\textbf{1P2P [$\%$]} & 73.1 & 82.9 & 86.8 & -20.2 & 44.2 & 93.0 & 27.8 & 52.4 & 65.1& 59.4 & 81.3 & 57.7 \\
\hline
\textbf{Var$\bm{(u)}$} & \multicolumn{12}{c}{} \\ \hline
\textbf{1P [$\%$]} & 24.1 & 36.0 & 35.6 & 36.5 & -15.0 & -12.8 & 44.2 & 5.1 & -12.8 & -72.7 & 26.0 & -3.8 \\ \hline
\textbf{1P2P [$\%$]} & -29.0 & 25.8 & -2.4 & 70.4 & 61.4 & -32.9 & 29.0 & 43.7 & 10.7 & 51.4 & 29.7 & -4.6 \\ \hline
\end{tabular}
\end{minipage}
\end{table}

Similar results are obtained for the experiments at different wind speeds and with other grid modes active. All these results are summarized in Table \ref{tab:results}. This table also shows the performance of SPRC when it only targets the 1P loads. 

Notice that out of the 12 experiments, SPRC outperforms CIPC in 10. Only in one case (Static 45$^{\circ}$ with 4$\,$m/s), SPRC is unable to reduce the variance of the loads. On average, SPRC for 1P and 2P achieves a reduction of load variance of 59$\,\%$, whereas conventional IPC leads to an average reduction of 49$\,\%$. Furthermore, on average, the variance of the pitch signals is 21$\,\%$ lower for SPRC compared to CIPC, indicating that the performance improvement does not come at the cost of a higher actuator duty.

Based on these results, it can be said that SPRC is able to reduce blade loads in more realistic high turbulent wind conditions. Next, the performance of SPRC in changing operating conditions will be discussed.

\subsection{Changing operating conditions}

The authors of \cite{SPRC2} show that SPRC is able to adapt to changing operating conditions in a laminar wind flow. In this section, it will be shown that similar results can be obtained in more turbulent wind conditions. Experiments are conducted where either the collective pitch angles or the wind speed is changed during operation. The performance of SPRC in these changing conditions will be evaluated.

\enlargethispage{-13.5\baselineskip}

With adaptive SPRC, the system parameters are being identified continuously: as defined in \eqref{eq:markovest}, new values of the Markov parameters are determined at every time instant $k$. Due to this feature, the algorithm is able to quickly adapt to changing operating conditions. As new measurements show a change in behavior, the system parameters will be changed accordingly. 

\begin{figure}[t]
\centering
\includegraphics[width=0.48\textwidth]{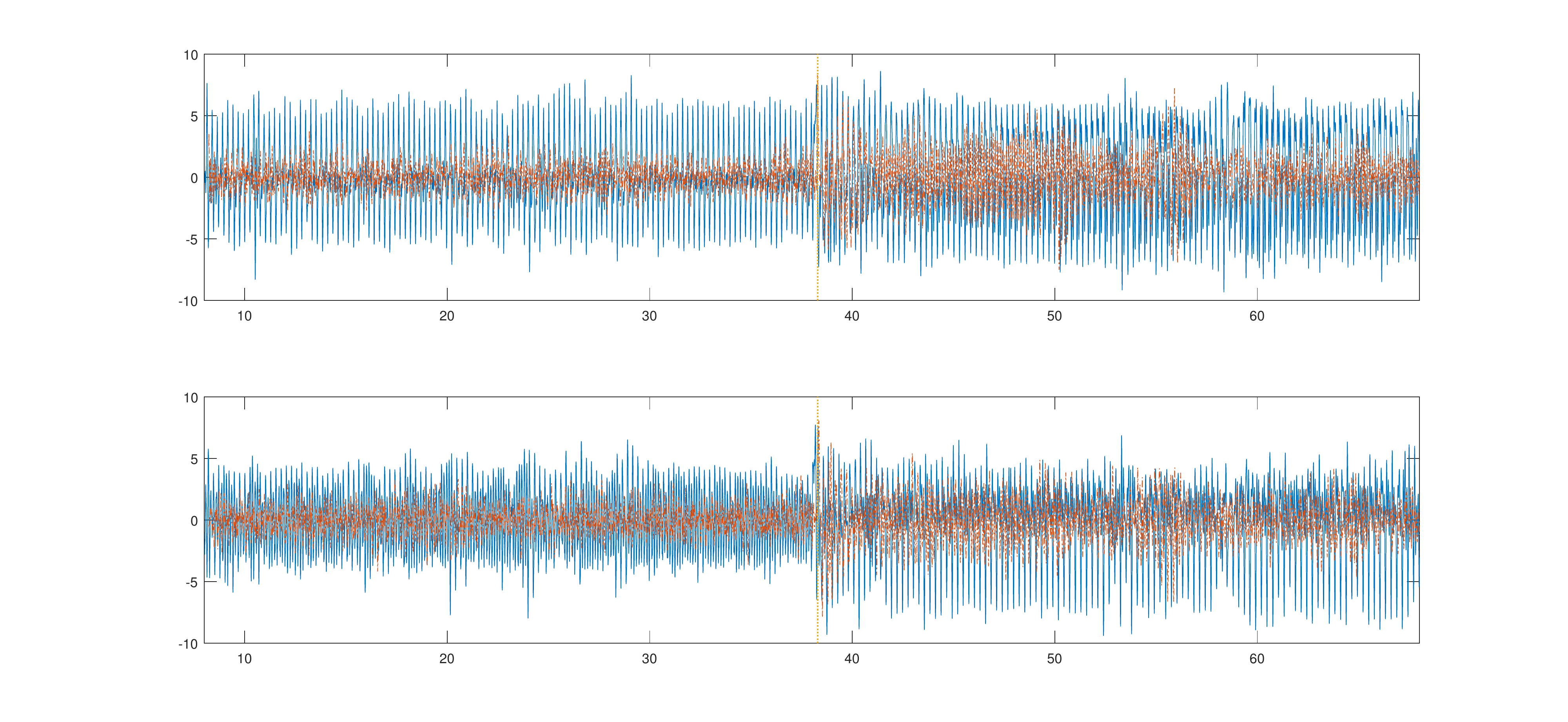}
\caption{Adaptive SPRC ({\color{red}-$\cdot$-$\cdot$-$\cdot$-}) versus no control ({\color{cyan}------}) for a change in collective pitch angle (from 2 to 10$^{\circ}$). Shown are the blade loads over the time of the experiments, as well as the pitch angles. The vertical line indicates the moment the collective pitch angles are changed.}
\label{fig:adaptpitch}
\end{figure}

\begin{figure}[h]
\centering
\includegraphics[width=0.48\textwidth]{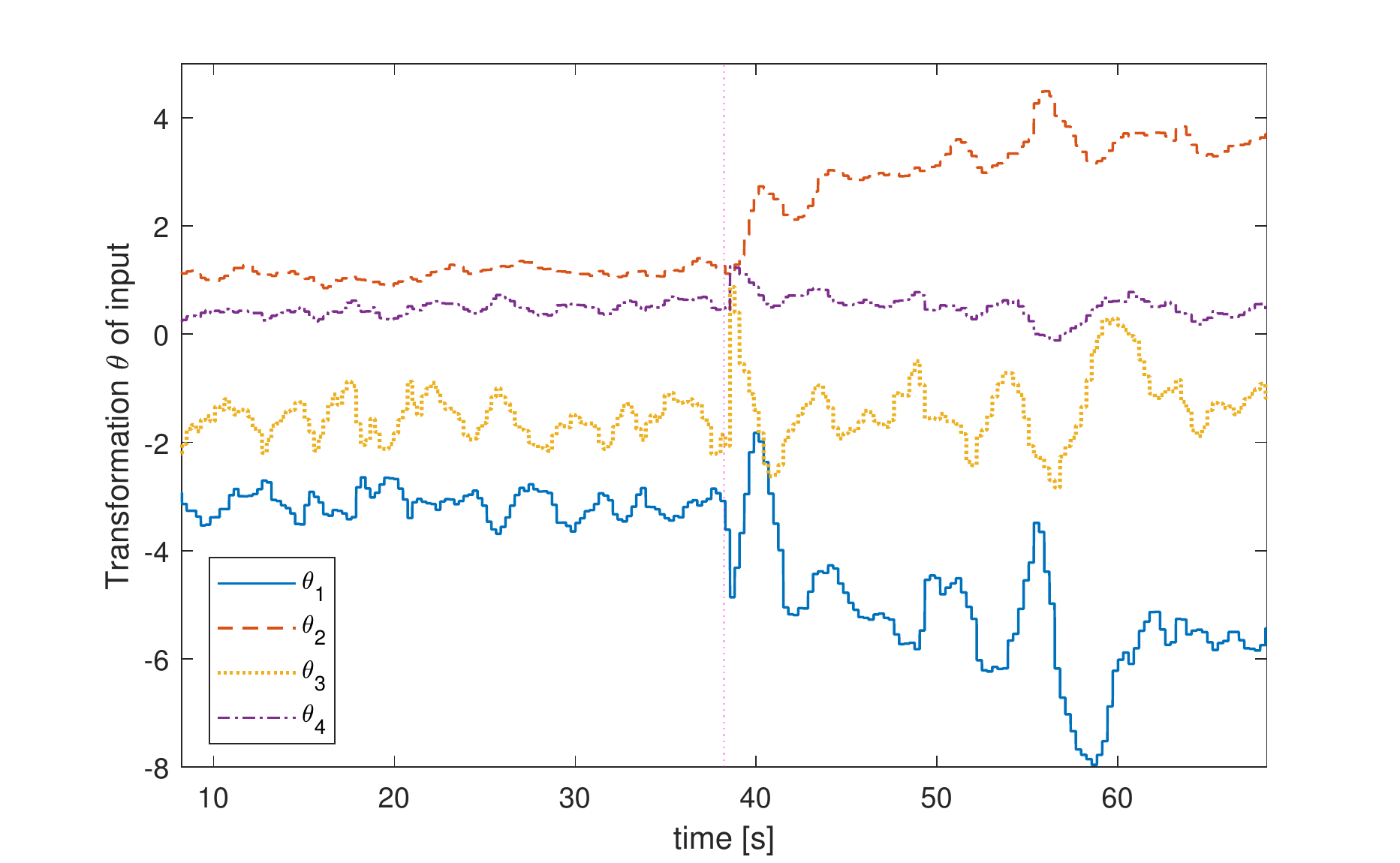}
\caption{The values of $\theta$ for blade 1, see \eqref{eq:ukdef}. The vertical line indicates the moment the collective pitch angles are changed.}
\label{fig:adapttheta}
\end{figure}

\begin{figure}[h]
\centering
\includegraphics[width=0.48\textwidth]{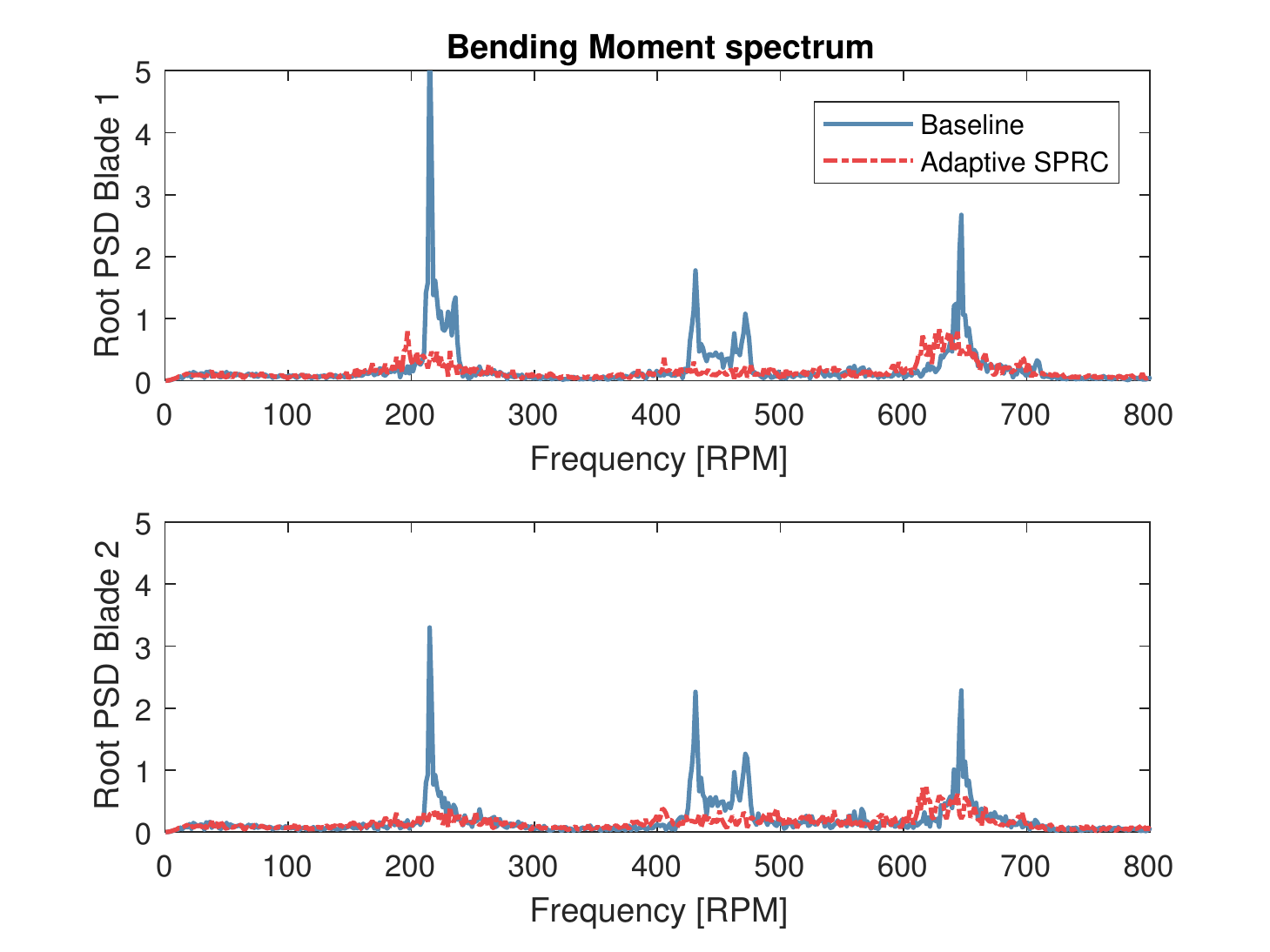}
\caption{Power spectrum of adaptive SPRC ({\color{red}- - -}) versus no control ({\color{cyan}------}) for a change in collective pitch angle (from 2 to 10$^{\circ}$). Notice two different peaks at each frequency due to the collective pitch change.}
\label{fig:adaptfd}
\end{figure}

The first adaptive experiment was conducted with a wind speed of 4.2$\,$m/s using the Gusts grid protocol. Figure \ref{fig:adaptpitch} shows the effect of changing the collective pitch during operation, resulting in a decrease of rotor speed from 240 to 210$\,$rpm. The blade loads and pitch angles for both blades are shown for SPRC and the baseline case of no control. This figure shows that the loads are again reduced after a short increase in blade loads when the pitch is changed at approximately 40 seconds. 

The amplitudes $\theta$ of the sinusoids that determine the pitch signal of blade 1 are shown in Figure~\ref{fig:adapttheta}. Here, it can be seen that the control input is quickly changed after the change of operating conditions. The oscillations in Figure~\ref{fig:adapttheta} show that the algorithm converges to the optimal values in approximately 15 seconds, and subsequently it can be seen in Figure~\ref{fig:adaptpitch} that the loads are reduced at the end of the experiment.

Figure \ref{fig:adaptfd} shows the power spectral density of the loads. As the rotor speed is changed due to the altered collective pitch, two peaks are visible at each harmonic. This figure demonstrates that SPRC significantly reduces the 1P and 2P blade loads of the turbine even when operating conditions are altered.

In the second experiment, the effect of a change in wind speed on the effectiveness of SPRC will be shown. This experiment is conducted using the static 45$^{\circ}$ grid protocol. During the experiment, the wind speed is increased from 4.5$\,$m/s to 5$\,$m/s, while the collective pitch stays constant at 2$^{\circ}$. This results in a significant increase in rotor speed: from approximately 200$\,$rpm to 240$\,$rpm.

\begin{figure}[t]
\centering
\includegraphics[width=0.48\textwidth]{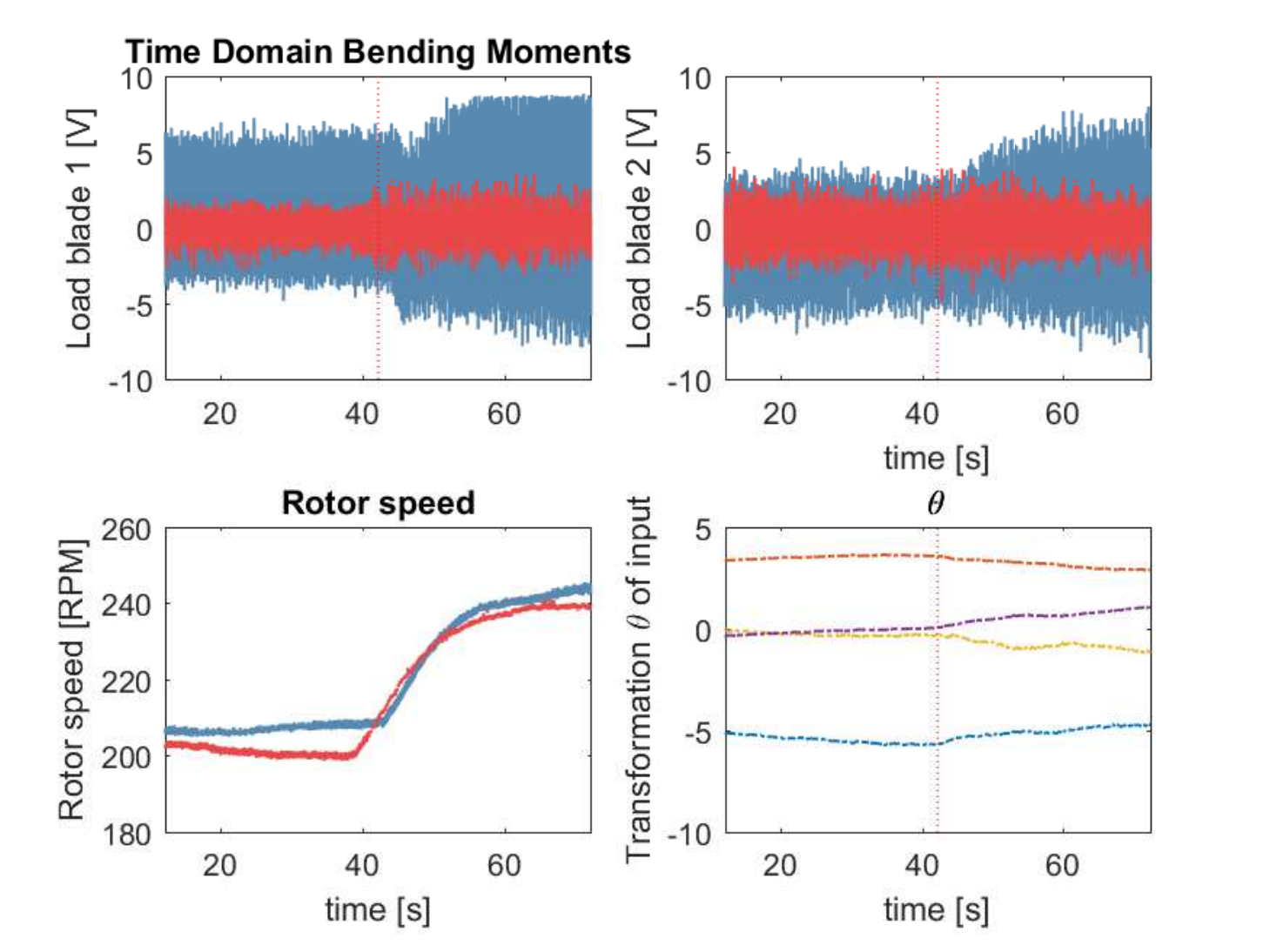}
\caption{Adaptive SPRC ({\color{red}-$\cdot$-$\cdot$-$\cdot$-}) versus no control ({\color{cyan}------}) for a wind speed change from 4.5$\,$m/s to 5$\,$m/s after approximately 40 seconds. In these experiments, the static 45$^{\circ}$ grid protocol was used. The upper figures show the loads of blade 1 and 2 respectively, the lower left figure the rotor speed, and the lower right figure the values of input $\theta$ for blade 1.}
\label{fig:adaptv}
\end{figure}

The results of this experiment are shown in Figure~\ref{fig:adaptv}. Comparing with Figure~\ref{fig:adaptpitch} shows that a changing wind speed has a smaller effect on the performance of SPRC than chaging the collective pitch. As can be seen in the bottom right figure, the control input only changes marginally after the wind speed is increased. The upper figures show that with SPRC, the loads barely increase when the wind speed increases, even though in the baseline case there is a substantial increase.

The two example cases in this section show that adaptive SPRC, where the system parameters are updated using online subspace identification, is able to quickly adjust the optimal RC when circumstances alter, resulting in a varying rotor speed, even in realistic turbulence conditions. 

\begin{figure}[t]
\centering
\includegraphics[width=0.48\textwidth]{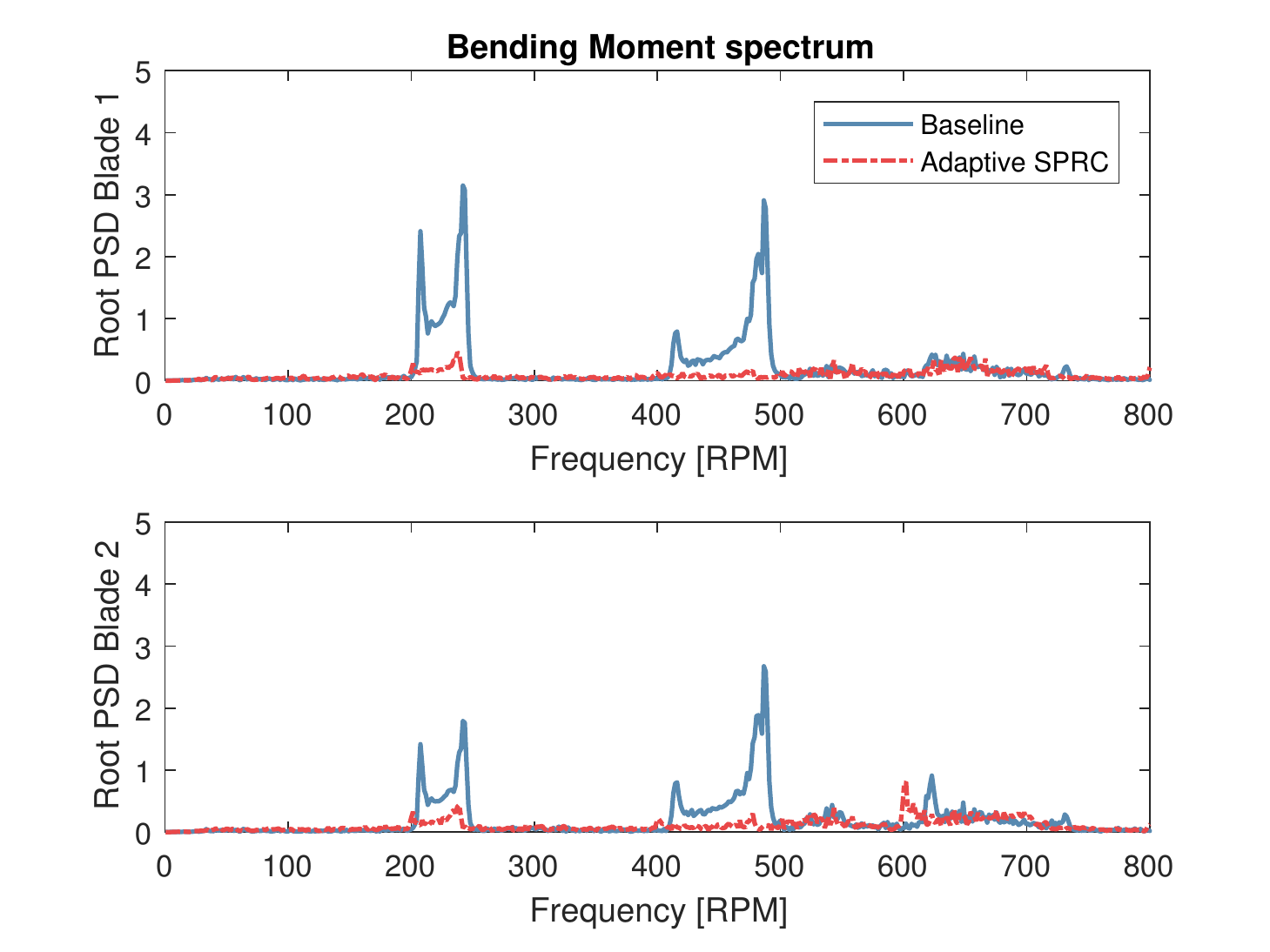}
\caption{The PSD of the loads of blade 1 (top) and 2 (bottom) using adaptive SPRC ({\color{red}-$\cdot$-$\cdot$-$\cdot$-}) versus no control ({\color{cyan}------}) for a wind speed change from 4.5$\,$m/s to 5$\,$m/s using the static 45$^{\circ}$ grid protocol.}
\label{fig:adaptvfd}
\end{figure}

\section{Conclusions}\label{sec:conclusions}

In this paper, a series of unique wind tunnel experiments have been conducted by implementing a large active grid in an open jet wind tunnel. It is shown that with this active grid, it is possible to produce wind conditions similar to real world conditions. Furthermore, the active grid creates reproducible wind conditions, which enables a fair evaluation of different control methodologies.

Using a two-bladed scaled wind turbine, the effectiveness of SPRC for individual pitch control in high-turbulent wind conditions is evaluated. Dedicated changes were made to the SPRC algorithm to ensure performance in the case of a varying rotor speed. The results of the wind tunnel experiments show that it is possible to reduce the variance of the blade loads significantly using SPRC under realistic high-turbulent wind conditions. This is achieved by specifically targetting the 1P and 2P loads on the blades using basis functions.

A comparison with conventional IPC shows that overall, SPRC outperforms CIPC in both low and high turbulent experiments. SPRC shows better performance in both blade load reduction and pitch actuator duty cycle. Averaged over all the different experiments, SPRC achieves a reduction of the blade load the variance of 59$\,\%$, an improvement of 10$\,\%$ compared to conventional IPC. Furthermore, the variance of the pitch angles is on average 21$\,\%$ lower than with conventional IPC. It can therefore be concluced that the SPRC algorithm is successfull at targetting only the relevant disturbances, and subsequently the load reduction is not obtained at the cost of a higher actuator duty cycle. This is supported by the power spectra of the blade loads, which shows a considerable reduction of the loads at the 1P and 2P frequencies.  

Finally, this paper shows that adaptive SPRC is able to handle changes in operating conditions, resulting in a varying rotor speed, even in high-turbulent wind conditions. Changes in pitch angles and in wind speed were applied, and in both cases, the algorithm quickly converges to a new optimum, maintaining performance. 

In conclusion, this paper demonstrates for the first time the performance of a data-driven repetitive individual pitch control algorithm under realistic wind conditions. Based on the results shown here, it can be concluded that SPRC is a very promising control methodology to achieve a load reduction of turbine blades.

\section*{Acknowledgments}

This work was supported by the INNWIND.EU Project, an EU Consortium with Academic and Industrial Partnership for Innovations in Wind Energy [grant agreement No.308974]. The authors would in addition like to thank C.J. Slinksman and Ing. W.J.M. van Geest for their help with the wind turbine setup.


\section*{References}

\begin{thebibliography}{00}


\bibitem[Barlas and Van Kuik, 2010]{review}
Barlas, Thanasis K. and Gijs A.M. van Kuik. "Review of state of the art in smart rotor control research for wind turbines." Progress in Aerospace Sciences 46.1 (2010): 1-27.
\bibitem[Bossanyi, 2000]{lqg1}
Bossanyi, Ervin A. "The design of closed loop controllers for wind turbines." Wind Energy 3.3 (2000): 149-163
\bibitem[Bossanyi, 2003]{ipc1}
Bossanyi, Ervin A. "Individual blade pitch control for load reduction." Wind Energy 6.2 (2003): 119-128
\bibitem[Bossanyi, 2003]{ipc2}
Bossanyi, Ervin A. "Further load reductions with individual pitch control." Wind Energy 8.4 (2005): 481-485
\bibitem[Bir, 2008]{coleman}
Bir, Gunjit. "Multiblade coordinate transformation and its application to wind turbine analysis." ASME Wind Energy Symposium. 2008.
\bibitem[Chiuso, 2007]{chiuso}
Chiuso, Alessandro. "The role of vector autoregressive modeling in predictor-based subspace identification." Automatica 43.6 (2007): 1034-1048
\bibitem[Dong and Verhaegen, 2008]{stability}
Dong, Jianfei and Verhaegen, Michel. "On the equivalence of closed-loop subspace predictive control with LQG." 47th IEEE Conference on Decision and Control (CDC) (2008): 4085-4090
\bibitem[Frederik et al., 2018]{ikke}
Frederik, Joeri, Lars Kr{\"o}ger, Jan-Willem van Wingerden, Joachim Peinke and Michael H{\"o}lling. "Data-driven individual pitch control: wind tunnel experiments under turbulent wind conditions." Proceedings of The Science of Making Torque from Wind (TORQUE) (2018)
\bibitem[Gustafsson, 2000]{gustafsson}
Gustafsson, Fredrik. "Adaptive filtering and change detection." New York: Wiley (2000)
\bibitem[Hallouzi et al., 2006]{hallouzi}
Hallouzi, Redouane, Michel Verhaegen, Robert Babuska and Stoyan Kanev. "Model weight and state estimation for multiple model systems applied to fault detection and identification." IFAC Proceedings Volumes 39.1 (2006): 648-653.
\bibitem[Hei{\ss}elmann et al., 2016]{heisselmann} 
Hei{\ss}elmann, Hendrik, Joachim Peinke, and Michael H\"{o}lling. "Experimental airfoil characterization under tailored turbulent conditions." J. Phys.: Conf. Ser. 753 (2016):  072020
\bibitem[Houtzager et al., 2013]{houtzager}
Houtzager, Ivo, Jan-Willem van Wingerden, and Michel Verhaegen. "Rejection of periodic wind disturbances on a smart rotor test section using lifted repetitive control." IEEE Transactions on Control Systems Technology 21.2 (2013): 347-359
\bibitem[Knebel, 2011]{knebel}
Knebel, Pascal, Achim Kittel, and Joachim Peinke. " Atmospheric wind field conditions generated by active grids." Exp. Fluids 51 (2011): 471–81
\bibitem[Kolmogorov, 1941]{kolmogorov}
Kolmogorov, A. N. "The local structure of turbulence in an incompressible viscous fluid for very high Reynolds number." Dokl. Akad. Nauk. SSSR (1941)
\bibitem[Kr\"{o}ger, 2018]{kroeger}
Kr\"{o}ger, Lars, Joeri Frederik, Jan-Willem van Wingerden, Joachim Peinke and Michael H\"{o}lling. " Validation of active control algorithms on a model wind turbine under controlled turbulent inflow conditions using an active grid." Proceedings of The Science of Making Torque from Wind (TORQUE) (2018)
\bibitem[Makita, 1991]{makita}
Makita, Hideharu. "Realization of a large-scale turbulence field in a small wind tunnel." Fluid Dynamics Research 8 (1991): 53-64
\bibitem[Mathworks, 2015a]{matlab}
Mathworks MATLAB-Simulink version R2015b (32-bit) (Released 2015). URL http://www.mathworks.com
\bibitem[Mathworks, 2015b]{slrt}
Mathworks Simulink Real-Time Documentation (2015). URL https://nl.mathworks.com/help/xpc/index.html
\bibitem[Mydlarski, 2017]{Mydlarski}
Mydlarski, Laurent. "A turbulent quarter century of active grids: from Makita (1991) to the present." Fluid Dynamics Research 49 (2017): 061401
\bibitem[Navalkar et al., 2014]{SPRC1}
Navalkar, Sachin T., Jan-Willem van Wingerden, Edwin van Solingen, Tom Oomen, Edwin Pasterkamp and Gijs van Kuik. "Subspace predictive repetitive control to mitigate periodic loads on large scale wind turbines." Mechatronics 24 (2014): 916-925
\bibitem[Navalkar et al., 2015]{SPRC2}
Navalkar, Sachin, Edwin van Solingen, and Jan-Willem van Wingerden. "Wind tunnel testing of subspace predictive repetitive control for variable pitch wind turbines." IEEE Transactions on Control Systems Technology 23.6 (2015): 2101-2116
\bibitem[Navalkar et al., 2016]{flaps}
Navalkar, Sachin, Lars Bernhammer, Jurij Sodja, Edwin van Solingen, Gijs van Kuik and Jan-Willem van Wingerden. "Wind tunnel tests with combined pitch and free-floating flap control: data-driven iterative feedforward controller tuning." Wind Energy Science 1.2 (2016): 205-220
\bibitem[Selvam et al., 2008]{ipc3}
Selvam, Kausihan, Stoyan Kanev, Jan-Willem van Wingerden, Tim van Engelen and Michel Verhaegen. "Feedback-feedforward individual pitch control for wind turbine load reduction." International Journal of Robust and Nonlinear Control 19.1 (2009): 72-91
\bibitem[Van Kuik and Peinke, 2016]{vkuik}
Van Kuik, Gijs and Joachim Peinke. "Long-term research challenges in wind energy - a research agenda by the european academy of wind energy." Wind Energy Science 1 (2016): 1-39.
\bibitem[Van Solingen et al., 2014]{edwin}
Van Solingen, Edwin, Sachin Navalkar and Jan-Willem van Wingerden. "Experimental wind tunnel testing of linear individual pitch control for two-bladed wind turbines." Journal of Physics: Conference Series Vol. 524, No. 1 (2014): 012056
\bibitem[Van Solingen et al., 2015]{edwin2}
Solingen, Edwin and Jan-Willem van Wingerden. "Linear individual pitch control design for two‐bladed wind turbines." Wind Energy 18.4 (2015): 677-697.
\bibitem[Van der Veen et al., 2013a]{vdveen}
Van der Veen, Gijs, Jan-Willem van Wingerden, and Michel Verhaegen. "Global identification of wind turbines using a Hammerstein identification method." IEEE transactions on control systems technology 21.4 (2013): 1471-1478
\bibitem[Van der Veen et al., 2013b]{vdveen2}
Van der Veen, Gijs, Jan-Willem van Wingerden, Marco Bergamasco, Marco Lovera and Michel Verhaegen. "Closed-loop subspace identification methods: an overview." IET Control Theory and Applications 7.10 (2013): 1339-1358
\bibitem[Verhaegen and Verdult, 2007]{verhaegen}
Verhaegen, Michel, and Vincent Verdult. "Filtering and system identification: a least squares approach." Cambridge university press, 2007

\end{thebibliography}
\end{document}